\documentclass[11pt]{article}
\usepackage{geometry}                % See geometry.pdf to learn the layout options. There are lots.
\usepackage{graphicx}
\usepackage{amssymb}
\usepackage{epstopdf}
\usepackage{amsthm}
\usepackage{color}
\author{Seth S. Cottrell\\Department of Physics and Astronomy\\Hunter College of the City University of New York\\New York, NY 10065}

\newtheorem{thm}{Theorem}[section]

\title{A Simple Method for Finding the Scattering Coefficients of Quantum Graphs}
\begin{document}
\maketitle

\pagebreak

\section*{Abstract}

Quantum walks are roughly analogous to classical random walks, and like classical walks they have been used to find new (quantum) algorithms. When studying the behavior of large graphs or combinations of graphs it is useful to find the response of a subgraph to signals of different frequencies.  In so doing we can replace an entire subgraph with a single vertex with frequency dependent scattering coefficients.

In this paper a simple technique for quickly finding the scattering coefficients of any quantum graph will be presented.  These scattering coefficients can be expressed entirely in terms of the characteristic polynomial of the graph's time step operator.  Moreover, with these in hand we can easily derive the ``impulse response" which is the key to predicting the response of a graph to any signal.  This gives us a powerful set of tools for rapidly understanding the behavior of graphs or for reducing a large graph into its constituent subgraphs regardless of how they are connected.

\pagebreak

\begin{tableofcontents}

\end{tableofcontents}

\pagebreak

\section{Introduction}

In classical computer science random walks have proven to be a useful tool for understanding and developing new algorithms and techniques.  The same has been true for quantum walks \cite{reitzner} and quantum computers.  A number  of these quantum walks have already been experimentally implemented, some using trapped ions \cite{ScMaScGlEnHuSc09,KaFoChStAlMeWi09} and others using photons in optical networks \cite{PeLaPoSoMoSi08}-\cite{schreiber}.  The goal of this paper is to provide a set of powerful and computationally cheap tools for rapidly understanding the behavior of graphs in terms of scattering signals.

A graph is a set of vertices with edges connecting those vertices.  In a classical random walk we imagine a particle that inhabits a vertex and has some probability of moving to connected vertices.  In a discrete random walk time is divided into integer steps and the probability of the particle jumping to an adjacent vertex in a given step is described by a stochastic matrix.  Like classical walks, the time can be either continuous \cite{farhi} or advance in discrete steps \cite{davidovich,vazirani}.  But unlike classical walks, with quantum walks the time step operator is unitary rather than stochastic.

In a classical walk the Hilbert space is composed entirely of the set of vertices, but in a quantum walk that isn't sufficient.  In a nutshell, there isn't enough information in a vertex state alone for time-reversibility (an essential characteristic of unitary processes) because at the very least the particle needs to ``know" where it was in the previous time step.  In this paper we'll use ``edge states" to more elegantly encode this information \cite{hillery0}.  For example, the edge state $|A,B\rangle$ is the state on the edge between vertices $A$ and $B$ that points from $A$ to $B$.  This is exactly equivalent to a particle on vertex $B$ that was previously on vertex $A$.  In this edge state formalism each vertex hosts a unitary operator that takes all of the incoming states and maps them to outgoing states.

Those already familiar with quantum walks will probably be more familiar with the ``coin space" \cite{reitzner} formalism, which is more directly analogous to classical random walks; the Hilbert space of discrete time coined quantum walks is the tensor product of the position space and the coin space \cite{Kollar}.  That is, each vertex is amended with an ancillary ``coin space".  The coin keeps track of where the particle previously was, and for this reason the dimension of the coin space associated with a vertex is always greater than or equal to the degree of that vertex.

Finally and most importantly, in a quantum walk the particle is in a superposition of states described by probability amplitudes (as opposed to probabilities).  When a position measurement is made the probability of finding a particle in the state $|\psi\rangle$ on the edge $|e\rangle$ is $|\langle e|\psi\rangle|^2$, in adherence to Born's rule.

Several previous studies have investigated scattering on quantum graphs.  Similar to the formality found in this paper, semi-infinite lines (``runways" of attached edges) are attached to a graph.  On these runways the time step operator passes the particle from one edge to the next sequentially, either toward or away from a particular vertex in the graph, and in this way the particle enters and exits the graph.  In \cite{farhi} scattering theory was used to show that tree graphs could be used to solve some kinds of decision problems using continuous time walks.  A discrete time scattering theory approach was fleshed out in \cite{feldman} where the connection between the connection between the number of steps to get through a graph and the transmission amplitude was found, as well as some more general results on the reflection and transmission amplitudes for a graph.  In \cite{modwalks} it was shown how the scattering matrices for subgraphs could be used to construct the scattering matrix for the overall graph.

In this paper we show that the semi-infinite runways of edges can be replaced with a single edge.  This allows us to analyze problems using strictly finite graphs, removing the issue of non-normalizable states.  More importantly, we show that the scattering coefficients, as well as the response to any incoming states, can be described entirely and succinctly by the characteristic polynomial of that finite graph's time step operator.  The time step operator is dictated by the structure of the graph, so this allows us to immediately see the relationship between the scattering coefficients of a graph and the structure of that graph.

Specifically, we find that the scattering coefficients of a graph are determined by the eigenstates and eigenvalues of the time step operator (which is now a finite matrix).  Fortunately, the techniques described in this paper do not at any time require these quantities to be calculated; instead we find that the characteristic polynomial itself is all that is needed.

\vspace{5mm}

In section 2 we consider the case of a single runway.  The problem and the exact definition of the effective reflection coefficient are defined rigorously.  Having only a single runway leads to some surprisingly compact results that are explored here.  We then use the frequency-dependent effective reflection coefficient to derive the graph's impulse response.  With this in hand we are able to rapidly calculate (with a convolution) the response to any arbitrary signal.

In section 3 we address the challenges of using a reflected signal to gain information about a graph and a theorem is proven which describes this difficulty explicitly.  Here we learn the relationship between the eigenvalues of the time step operator and the length of a signal on the runway necessary to detect the effect of those eigenvalues.  This is an important tool for understanding the computational time of algorithms.

In section 4 we explore the case of a graph attached to multiple runways, and the effective scattering coefficient between them.  A very powerful theorem is derived that allows for the rapid and simultaneous calculation of each of these coefficients in terms of the resolvent.

Finally, in the appendix (the second half of this paper) there are a series of examples that put all of the theorems and techniques described in this paper to use, demonstrating how simple they are in practice.

\pagebreak

\section{Basic Framework}

The situation in question is an arbitrary graph, $G$, attached to an infinite ``runway" of edges.  The vertices on the runway are labeled 0, 1, 2, ... where 0 is a given vertex of $G$.  We define the unitary time step operator on the runway as the one that passively moves each edge state one step.  I.e., ${\bf U}|j,j+1\rangle = |j+1,j+2\rangle$ and ${\bf U}|j+1, j\rangle = |j,j-1\rangle$.  The behavior of ${\bf U}$ in the graph $G$ is not specified here.

Our goal is to replace the graph $G$ with a {\it single} vertex and to encode all of its behavior into that one vertex.  A set of constant reflection/scattering coefficients doesn't contain nearly enough information to simulate the behavior of a complicated graph, but if we allow them to be frequency-dependent, then this goal is attainable.

The frequency-dependent scattering coefficient is defined such that it replaces the entire graph with a single reflection coefficient at vertex zero with value $S(\lambda)$.  When dealing with multiple runways attached to the same graph we can consider them one at a time, since the time step operator is linear.

\begin{figure}[h!]
\centering
\includegraphics[width=2.8in]{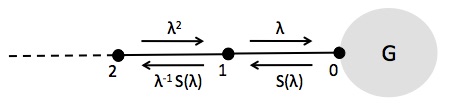}
\caption{A signal that advances by $\lambda$ every time step produces a ``reflection" that is shifted by some phase.  The scattering coefficient is defined such that the graph can be replaced with a single vertex that reflects with $S(\lambda)$.}
\end{figure}

In this section we'll first consider a graph with one connected runway.  It will be demonstrated that we can understand the response of a graph connected to an infinite runway by looking at the characteristic polynomial of the graph alone.

We define an operator, ${\bf U}$, on $G$ and the runway such that ${\bf U}|1,0\rangle = |in\rangle$ and ${\bf U}|out\rangle = |0,1\rangle$, and $\forall n>0$, ${\bf U}|n-1,n\rangle = |n,n+1\rangle$ and ${\bf U} |n+1,n\rangle = |n,n-1\rangle$.

\vspace{5mm}

An incoming pure momentum state takes the form $\sum_{j=0}^\infty \lambda^{j+1} |j+1,j\rangle$.  If vertex 0 is completely reflective with reflection coefficient $r$, then the response is $\sum_{j=0}^\infty r\lambda^{n-j} |j,j+1\rangle$ (see figure 1).  Clearly,

\begin{equation}
|\Psi\rangle = \sum_{j=0}^\infty \lambda^{j+1} |j+1,j\rangle + \sum_{j=0}^\infty r\lambda^{-j} |j,j+1\rangle
\end{equation}

is an eigenstate with eigenvalue $\lambda$.

In this form it's easier to see how this is a signal and reflection.  After $n$ time steps this will take the form  $\lambda^n|\Psi\rangle = \sum_{j=0}^\infty \lambda^{n+j+1} |j+1,j\rangle + \sum_{j=0}^\infty r\lambda^{n-j} |j,j+1\rangle$, and after $n+1$ time steps the state is $\lambda^{n+1}|\Psi\rangle = \sum_{j=0}^\infty \lambda^{n+j+2} |j+1,j\rangle + \sum_{j=0}^\infty r\lambda^{n+1-j} |j,j+1\rangle$.  The coefficient of $|0,1\rangle$ is $r$ times the coefficient of $|1,0\rangle$ in the {\it previous} time step, which is exactly as it should be.

If instead of a single simply-reflecting vertex there is a graph attached to vertex 0, then the eigenstate now takes the form

\begin{equation}\label{psi}
|\Psi\rangle = \sum_{j=0}^\infty \lambda^{j+1} |j+1,j\rangle + \sum_{j=0}^\infty S(\lambda)\lambda^{-j} |j,j+1\rangle + |G\rangle
\end{equation}

where $|G\rangle$ is the component of the eigenstate contained in $G$.  In this way we can define an ``effective reflection coefficient", $S(\lambda)$.  Unless $G$ is a single vertex, $S(\lambda)$ will be a non-constant function of $\lambda$.  In either case, equation \ref{psi} is a $\lambda$-eigenstate of the graph and runway.

\begin{figure}[h!]
\centering
\includegraphics[width=3.0in]{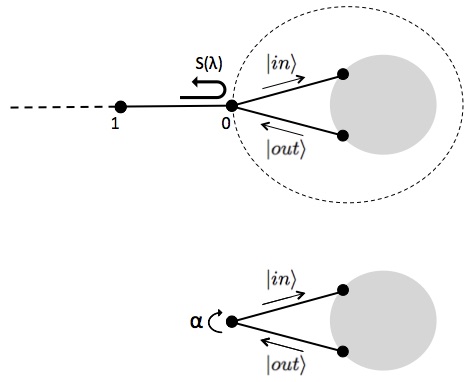}
\caption{${\bf U}$ and ${\bf U}_\alpha$.  By selecting the correct value of $\alpha$ we can produce a $\lambda$-eigenstate of ${\bf U}_\alpha$ that is identical to the $\lambda$-eigenstate of ${\bf U}$ on the edges they have in common (all of which are in $G$).}
\end{figure}

In order to find $S(\lambda)$ we create a new operator, ${\bf U}_\alpha$, that reflects back into the graph rather than transmitting into or receiving from the runway.  That is:

\begin{equation}
{\bf U}_\alpha |out\rangle = \alpha |in\rangle
\end{equation}

We will find that there is a simple relationship between $\alpha$ and $S(\lambda)$, and that we can determine the correct value of $\alpha$ by tuning it such that ${\bf U}_\alpha$ has $\lambda$ as an eigenvalue.

First we need to derive a few properties of the characteristic polynomial of ${\bf U}_\alpha$.

\vspace{5mm}

\begin{thm}\label{polynomial}
$C(z,\alpha) = \left|{\bf U}_\alpha - z{\bf I}\right| = b(z)(f(z) + \alpha g(z))$, where $f(z)$, $g(z)$, and $b(z)$ are polynomials in $z$.  $f(z)$ and $g(z)$ share no common roots, the roots of $b(z)$ sit on the unit circle, and the roots of $f(z)$ sit strictly within the unit circle.
\end{thm}

{\it Proof}

This is easy to immediately verify by inspection of the matrix ${\bf U}_\alpha - z{\bf I}$.  $\alpha$ appears once, so every term in the characteristic polynomial either contains an $\alpha$ or doesn't.  Clearly, the characteristic polynomial is affine in $\alpha$.

We can collect the terms with and without $\alpha$'s into two polynomials.  Trivially, those polynomials can be labeled $b(z)f(z)$ and $\alpha b(z)g(z)$, where $b(z)$ is the collection of all of the factors common to both polynomials.

Since the roots of $b(z)$ are independent of $\alpha$, and since ${\bf U}_\alpha$ can be unitary (when $|\alpha|=1$), the roots of $b(z)$ are eigenvalues of a unitary matrix and therefore have modulus 1.

Clearly, $f(z)=|{\bf U}_0-z{\bf I}|$, where ${\bf U}_0 := {\bf U}_\alpha\big|_{\alpha=0}$.  Define $|\Psi_0\rangle = a|out\rangle+|G\rangle$ to be a normalized eigenstate of ${\bf U}_0$ and $\eta$ to be a root of $f(z)$.  Since ${\bf U}_0|out\rangle = 0|in\rangle$, we know that $\langle in|\Psi_0\rangle = 0$.  When $|\alpha|=1$ we know that ${\bf U}_\alpha$ is unitary and therefore

$\begin{array}{ll}
|\eta|^2 = \langle\Psi_0|{\bf U}_0^\dagger {\bf U}_0|\Psi_0\rangle \\[2mm]
= \langle\Psi_0|\left({\bf U}_\alpha^\dagger - \alpha^*|out\rangle\langle in|\right)\left({\bf U}_\alpha - \alpha|in\rangle\langle out|\right)|\Psi_0\rangle \\[2mm]
= \langle\Psi_0|{\bf U}_\alpha^\dagger{\bf U}_\alpha|\Psi_0\rangle - 2Re\left[ \alpha^*\langle\Psi_0|out\rangle\langle in|{\bf U}_\alpha|\Psi_0\rangle \right] + |\alpha|^2\langle\Psi_0|out\rangle\langle in|in\rangle\langle out|\Psi_0\rangle \\[2mm]
= \langle\Psi_0|\Psi_0\rangle - 2Re\left[ a^*\alpha^*\langle in|{\bf U}_\alpha|\Psi_0\rangle \right] + |\alpha|^2|a|^2 \\[2mm]

= 1 - 2Re\left[ a^*\alpha^*\left(a\alpha\right) \right] + |\alpha|^2|a|^2 \\[2mm]
= 1 - |\alpha|^2|a|^2 \\[2mm]
=1 - |a|^2 \\[2mm]
<1
\end{array}$ 

If $a=0$, then $|\Psi_0\rangle$ is a bound eigenstate, and $\eta$ would actually be a root of $b(z)$.  Therefore all of the roots of $f(z)$ are inside the unit circle.

$\square$

\vspace{5mm}

In everything that follows $b(z)$ either doesn't play a roll, isn't relevant, or factors out.  So, it will be suppressed.  So far we've replaced one unknown variable, $S(\lambda)$, with another, $\alpha$, however this is a step forward because we can quickly find a closed solution for $\alpha$.

\vspace{5mm}

\begin{thm}\label{important1}
$S(\lambda) = \frac{1}{\alpha} = -\frac{g(\lambda)}{f(\lambda)}$.
\end{thm}

{\it Proof}

This is the essential trick of this paper.

The $\lambda$-eigenstate of ${\bf U}$ takes the form $|\Psi\rangle=\sum_{k=0}^\infty \lambda^{k+1}|k+1,k\rangle + S(\lambda)\sum_{k=0}^\infty \lambda^{-k}|k,k+1\rangle + \lambda S(\lambda)|out\rangle + |in\rangle + |G\rangle$.

The $\lambda$-eigenstate of ${\bf U}_\alpha$ takes the form $|\Psi_\alpha\rangle = \frac{\lambda}{\alpha}|out\rangle + |in\rangle + |G\rangle$.

Restricted to $G$, these two states are the same.  But whereas the coefficient of $|in\rangle$ in $|\Psi\rangle$ is dictated by the incoming signal on the runway, in $|\Psi_\alpha\rangle$ it's dictated by the coefficient of $|out\rangle$ and the value of $\alpha$.  By tuning $\alpha$ to the correct value we're ``feeding the output to the input" so that $\langle in|{\bf U}^n|\Psi\rangle = \langle in|{\bf U}_\alpha^n|\Psi_\alpha\rangle = \lambda^n$, $\forall n$.

Equating the coefficients of the $|out\rangle$ states in these two eigenstates we see immediately that $S(\lambda)=\frac{1}{\alpha}$.  We can then solve for $\alpha$ using the characteristic equation.  When $\lambda$ is an eigenvalue we have that $0=f(\lambda)+\alpha g(\lambda)$ and therefore $\alpha = -\frac{f(\lambda)}{g(\lambda)}$.  It follows that $S(\lambda)=-\frac{g(\lambda)}{f(\lambda)}$.  Note that $S(z)$ is a meromorphic function of $z$.

$\square$

\vspace{5mm}

To reiterate and make clear what $S(\lambda)$ is: so long as the runway and $G$ are in the $\lambda$-eigenstate, we can replace the $G$ with a reflection coefficient, $S(\lambda)$, at vertex 0.  In this way we can describe the graph's reaction to an infinite signal with frequency $\lambda=e^{i\theta}$.

\subsection{Particulars for a Single Runway}

In the case of a single input and output ${\bf U}_\alpha$ is unitary for $|\alpha|=1$.  This unitarity has a lot of consequences, but in particular $S(z)$ takes the following form:

\begin{equation}\label{compliment}
S(z) = \frac{1}{\alpha} = -\frac{g(z)}{f(z)} = -\frac{g_0}{z^{s} } \prod_j \frac{1-z\eta_j^*}{z - \eta_j}
\end{equation}

where $|\eta_j|<1$, $\forall j$.

\vspace{5mm}

In this section we'll explore the special case of a single runway.

\vspace{5mm}

\begin{thm}
$C(z,\alpha) = g_0\alpha z^dC^*\left(\frac{1}{z}, \frac{1}{\alpha}\right), \quad \forall z,\forall\alpha \ne 0$ where $C^*$ indicates the coefficients are conjugated, $d$ is the degree of $C(z,\alpha)$, and $g_0$ is the constant term of $g(z)$.  Equivalently, $f(z) = z^s\prod_{j=1}^{d^\prime} \left(z-\eta_j\right), \quad g(z) = g_0z^df^*\left(\frac{1}{z}\right) = g_0\prod_{j=1}^{d^\prime} \left(1-z\eta_j^*\right)$, where $d^\prime+s=d$.
\end{thm}

{\it Proof}

In what follows assume that $|\alpha| = 1$.  This means that ${\bf U}_\alpha$ is unitary, and the roots of the associated characteristic polynomial, $C(z,\alpha)$, all have modulus 1.  While this proof will only consider $f(z)$ and $g(z)$, it works in exactly the same way for $b(z)f(z)$ and $b(z)g(z)$.

Let $C(z,\alpha) = \prod_{k=1}^d \left(z - \lambda_k\right) = \sum_{k=0}^d f_k z^k + \alpha\sum_{k=0}^d g_k z^k$.  Note that $f_d=1$ and $g_d = g_{d-1}=0$, since when the determinant was taken any term with $\alpha$ necessarily did not include 2 diagonal elements (2 powers of $z$).

\vspace{5mm}

$\begin{array}{ll}
f(\lambda_k) + \alpha g(\lambda_k) = 0 \\[2mm]
\Leftrightarrow C(\lambda_k,\alpha) = 0 \\[2mm]
\Leftrightarrow \left(C(\lambda_k, \alpha)\right)^* = 0 \\[2mm]
\Leftrightarrow C^*(\lambda_k^*, \alpha^*) = 0 \\[2mm]
\Leftrightarrow C^*\left(\frac{1}{\lambda_k}, \frac{1}{\alpha}\right) = 0 \\[2mm]
\Leftrightarrow 0 = f^*\left(\frac{1}{\lambda_k}\right) + \frac{1}{\alpha}g^*\left(\frac{1}{\lambda_k}\right) \\[2mm]
\end{array}$

Therefore, $C(z, \alpha)$ and $C^*\left(\frac{1}{z}, \frac{1}{\alpha}\right)$ have the same set of zeros.  It also follows that $\alpha z^d C^*\left(\frac{1}{z}, \frac{1}{\alpha}\right) = \alpha z^df^*\left(\frac{1}{z}\right) + z^dg^*\left(\frac{1}{z}\right)$ is a polynomial in $z$ and $\alpha$ which, again, has the same set of zeros.  Therefore, $\alpha z^d C^*\left(\frac{1}{z}, \frac{1}{\alpha}\right)$ and $C(z,\alpha)$ are proportional to each other.

$\begin{array}{ll}
bC(z,\alpha) \\[2mm]
=\alpha z^dC^*\left(\frac{1}{z},\frac{1}{\alpha}\right) \\[2mm]
= \alpha z^d\left[\sum_{k=0}^d f_k^* \frac{1}{z^k} + \frac{1}{\alpha}\sum_{k=0}^d g_k^* \frac{1}{z^k}\right] \\[2mm]
= \alpha\sum_{k=0}^d f_k^*z^{d-k} + \sum_{k=0}^d g_k^* z^{d-k} \\[2mm]
= \alpha\sum_{k=0}^d f_{d-k}^*z^{k} + \sum_{k=0}^d g_{d-k}^* z^{k} \\[2mm]
\Rightarrow \left\{\begin{array}{ll}
bf_k = g_{d-k}^* \\
bg_k = f_{d-k}^* \\
\end{array}\right.
\end{array}$

Now,

$f_d=1 \Rightarrow  b=g_0^*$

We now have that $C(z,\alpha) = g_0\alpha z^d C^*\left(\frac{1}{z}, \frac{1}{\alpha}\right)$.

Since the constant term in a characteristic equation is equal to the determinant, $1=|\alpha g_0|=|g_0|$, which means that $0$ is not a root of $g(z)$.  Because $g_d=g_{d-1}=0$ we have that $f_0=f_1=0$, which implies that $s\ge 2$

Keep in mind that the result above is merely a statement about the polynomial $C(z,\alpha)$.  It is true regardless of the value of $\alpha$.

\begin{equation}
C(z,\alpha) = g_0\alpha z^d C^*\left(\frac{1}{z},\frac{1}{\alpha}\right)
\end{equation}

in general, $\forall \alpha\ne0$.  Or equivalently,

\begin{eqnarray}
f(z) = g_0z^d g^*\left(\frac{1}{z}\right)\\
g(z) = g_0z^d f^*\left(\frac{1}{z}\right)
\end{eqnarray}

$f(z)$ and $g(z)$ are said to be ``reciprocal polynomials" of each other.

$\square$

\vspace{5mm}

We can say even more about the characteristic polynomial.  The zeros of $f(z)$ and $g(z)$ have a very particular behavior and relationship.

\vspace{5mm}

\begin{thm}
$C(z,\alpha) = \underbrace{z^s\prod_j \left(z-\eta_j\right)}_{f(z)} + \alpha\underbrace{g_0\prod_j \left(1-z\eta_j^*\right)}_{g(z)}$

where $0<|\eta_j|<1$, $\forall j$.
\end{thm}

{\it Proof}

We already know that $|\eta_j|<1$ from theorem \ref{polynomial} and that $f(z) = g_0z^d g^*\left(\frac{1}{z}\right)$ from the last theorem.  It follows that for $\eta_j\ne0$, $f(\eta_j)=0 \Leftrightarrow g\left(\frac{1}{\eta_j^*}\right) = 0$.  Therefore, if $f(z) = z^s\prod_j \left(z - \eta_j\right)$, then $g(z)\propto\prod_j \left(z - \frac{1}{\eta_j^*}\right)\propto\prod_j \left(1 - z\eta_j^*\right)$.  With $g_0$ the constant term in $g(z)$, we can write $g(z) = g_0 \prod_j \left(1 - z\eta_j^*\right)$.

Notice that $f(z)$ and $g(z)$ share a root if and only if $\eta_j = \frac{1}{\eta_j^*}$ or $|\eta_j| = 1$.  But this is exactly what we expect for the roots of $b(z)$.
$\square$

\vspace{5mm}

The above statement about the roots of $f(z)$ applies more generally; that is, it continues to apply when there are multiple inputs and outputs.

\vspace{5mm}

\begin{thm}\label{distinct}
When $|\alpha|=1$, the solutions of $C(z, \alpha)$ are distinct.
\end{thm}

{\it Proof}

When $|\alpha|=1$ we know that ${\bf U}_\alpha$ is unitary.  An immediate consequence of which is the fact that ${\bf U}_\alpha$ is diagonalizable and expressible as ${\bf U}_\alpha = \sum_\lambda \lambda{\bf P}_\lambda$, where ${\bf P}_\lambda$ is a projection operator onto the $\lambda$-eigenspace.  Each of these projections can be expressed in terms of the resolvent, which in turn can be written as a power series in $\alpha-\alpha_0$ near $\alpha_0$, where $\alpha_0$ is any arbitrary point on the unit circle.

This implies that the projection operators can likewise be expressed as a power series in $\alpha-\alpha_0$, and since  ${\bf P}_\lambda = |V_\lambda\rangle\langle V_\lambda|$, it follows that the eigenvectors share the same property.  Finally, since ${\bf U}_\alpha$ and $|V_\lambda\rangle$ are power series in $\alpha-\alpha_0$, and ${\bf U}_\alpha|V_\lambda\rangle = \lambda|V_\lambda\rangle$, we can see that the eigenvalues themselves, $\lambda$, are power series in $\alpha-\alpha_0$.

Now define $c_0(z)(z-\lambda_0)^t = f(z) + \alpha_0g(z)$.  Note that according to the last theorem $g(z)\ne0$ when $|z|=1$.

\vspace{5mm}

$\begin{array}{ll}
0 = f(\lambda) + \alpha g(\lambda) \\[2mm]
= f(\lambda) + \alpha_0 g(\lambda) + (\alpha-\alpha_0) g(\lambda) \\[2mm]
= c_0(\lambda)(\lambda-\lambda_0)^t + (\alpha-\alpha_0) g(\lambda) \\[2mm]
\Rightarrow (\lambda-\lambda_0)^t = - \frac{g(\lambda)}{c_0(\lambda)}(\alpha-\alpha_0)  \\[2mm]
\Rightarrow \lambda = \lambda_0 + O\left(\sqrt[t]{\alpha-\alpha_0}\right)  \\[2mm]
\end{array}$

However, since the eigenvalues are expressible as a power series in $\alpha-\alpha_0$, $t=1$.  Therefore, because $\alpha_0$ is arbitrary, the degree of any zero of $f(z) + \alpha g(z)$ is one when $|\alpha|=1$.

$\square$

\vspace{5mm}

In this proof it was important that $|\alpha|=1$ because it ensures that ${\bf U}_\alpha$ is unitary.  For a finite set of values of $\alpha$ (off of the unit circle) we find that $f(z)+\alpha g(z)$ can have higher degree roots, however at those points we find that ${\bf U}_\alpha$ is no longer diagonalizable and the degenerate eigenvalues correspond to {\it generalized} eigenvectors.

\vspace{5mm}

\begin{thm}\label{permute}
When $\alpha$ loops once around the unit circle the eigenvalues cyclicly permute one step.  That is, looping $\alpha$ changes $\lambda_j\to\lambda_{j+1}$ and $\lambda_d\to\lambda_1$, where $arg\left(\lambda_1\right) < arg\left(\lambda_2\right) < \cdots < arg\left(\lambda_d\right)$.
\end{thm}

We know that looping $\alpha$ once (returning it to its original value) can't change the spectrum of the eigenvalues, so the effect must be a permutation.  In addition, since the eigenvalues are always distinct for every value of $|\alpha|=1$, this permutation must be cyclic (the eigenvalues can't "slide past each other" on the unit circle).

So we know that looping $\alpha$ produces a permutation of the eigenvalues of the form $\lambda_j\to\lambda_{j+t}$ (where $\lambda_d\equiv\lambda_0$).  The only question that remains is the value of $t$.

Define $\lambda = e^{i\theta}$.  The eigenvalues satisfy

$\begin{array}{ll}
0 = f(\lambda) + \alpha g(\lambda) \\[2mm]
\Rightarrow  -f\left(e^{i\theta}\right) = \alpha g\left(e^{i\theta}\right) \\[2mm]
\Rightarrow  -e^{is\theta}\prod_{j=1}^{d^\prime} \left(e^{i\theta} - \eta_j\right) = \alpha g_0\prod_{j=1}^{d^\prime} \left(1 - e^{i\theta}\eta_j^*\right) \\[2mm]
\Rightarrow  -e^{is\theta}\prod_{j=1}^{d^\prime} \left(e^{i\theta} - \eta_j\right) = \alpha g_0\prod_{j=1}^{d^\prime} e^{i\theta} \left(e^{i\theta} - \eta_j\right)^* \\[2mm]
\Rightarrow  -e^{is\theta}\prod_{j=1}^{d^\prime} \left(e^{i\theta} - \eta_j\right) = \alpha g_0 e^{id^\prime \theta} \prod_{j=1}^{d^\prime} \left(e^{i\theta} - \eta_j\right)^* \\[2mm]
\Rightarrow \alpha = -g_0^*e^{i(s-d^\prime)\theta}\prod_{j=1}^{d^\prime} \frac{\left(e^{i\theta} - \eta_j\right)}{\left(e^{i\theta} - \eta_j\right)^*} \\[2mm]
\Rightarrow \log(\alpha) = i\pi + \log(g_0) + i(s-d^\prime)\theta + \sum_{j=1}^{d^\prime} \log\left( \frac{\left(e^{i\theta} - \eta_j\right)}{\left(e^{i\theta} - \eta_j\right)^*}\right) \\[2mm]
\Rightarrow i\arg(\alpha) = i\pi + i\arg(g_0) + i(s-d^\prime)\theta + \sum_{j=1}^{d^\prime} i2\arg\left(e^{i\theta} - \eta_j\right) \\[2mm]
\end{array}$

We now have a relation between the zeros of $f(z)$ and the phase of $\alpha$.

\begin{equation}\label{phase}
\arg(\alpha) = \pi + \arg(g_0) + (s-d^\prime)\theta + 2\sum_{j=1}^{d^\prime} \arg\left(e^{i\theta} - \eta_j\right)
\end{equation}

At this point we allow $\theta$ to smoothly increase by $2\pi$, then take the difference.  Since $|\eta_j|<1$, the angle between $e^{i\theta}$ and $\eta_j$ sweeps from $0$ to $2\pi$ monotonically.

$\begin{array}{ll}
\Rightarrow  \Delta arg\left(\alpha\right) = (s-{d^\prime})2\pi + 2\sum_{j=1}^{d^\prime} 2\pi \\[2mm]
\Rightarrow  \Delta arg\left(\alpha\right) = (s+{d^\prime})2\pi \\[2mm]
\Rightarrow  \Delta arg\left(\alpha\right) = 2\pi d \\[2mm]
\end{array}$

Looping a given eigenvalue once around the unit circle causes $\alpha$ to loop $s+d^\prime = d$ times.  Looping an eigenvalue once is a permutation of the form $\lambda_{j}\to\lambda_{j+d} = \lambda_j$.  It follows that if looping $\alpha$ once produces a permutation of the form $\lambda_j\to\lambda_{j+t}$, then looping $\lambda_j$ means that $\alpha$ loops $\frac{d}{t}$ times.  But we know that looping an eigenvalue once requires $\alpha$ to loop $d$ times, and therefore $t=1$.

$\square$

\vspace{5mm}

\begin{thm}\label{unique}
Any eigenvalue $\lambda$, such that $|\lambda|=1$, can be induced by choosing the correct value of $\alpha$.  Moreover, this value is unique.
\end{thm}

In the last theorem it was shown that the eigenvalues, which are functions of $\alpha$, cyclicly permute when $\alpha$ loops around the unit circle.  These functions are continuous, so every value between these eigenvalues exist for some value of $\alpha$ as well.  Moreover, $arg(\lambda)$ is a strictly monotonic function of $arg(\alpha)$, and therefore the value of $\alpha$ is unique for a given eigenvalue.

This can be seen by first showing that $\frac{\partial}{\partial \theta} arg\left(e^{i\theta} - \eta\right) > \frac{1}{2}$ when $|\eta|<1$.  This can be proven by either using the inscribed angle theorem to establish a lower bound or by direct calculation.  It follows that 

$\begin{array}{ll}
arg\left(\alpha\right) = \pi + (s-d^\prime)\theta + i\log{(g_0)} + 2\sum_{j=1}^{d^\prime} arg\left(e^{i\theta} - \eta_j\right) \\[2mm]
\Rightarrow \frac{\partial}{\partial \theta} arg\left(\alpha\right) = (s-d^\prime) + 2\sum_{j=1}^{d^\prime} \frac{\partial}{\partial \theta} arg\left(e^{i\theta} - \eta_j\right) > (s-d^\prime) + 2d^\prime\left(\frac{1}{2}\right) = s \ge 0 \\[2mm]
\Rightarrow \frac{\partial}{\partial \theta} arg\left(\alpha\right) > 0 \\[2mm]
\end{array}$

Therefore $\arg(\alpha)$ and $\arg(\lambda)$ are strictly monotonic functions of each other.  This monotonicity ensures the uniqueness of $\alpha$ for a given value of $\lambda$ by making sure that $\arg(\alpha(\lambda))$ doesn't double back on itself.

$\square$

\subsection{Arbitrary Inputs for a Single Runway}

In the language of signal analysis, the last section was a derivation of the ``frequency response" of the graph $G$.  We can define the input $x[n]$ (output $y[n]$) as the amplitude on the state $|1,0\rangle$ (state $|0,1\rangle$) at time step $n$ (time step $n+1)$.  The input can be encoded onto the runway in an initial state of the form $\sum_{k=0}^\infty x[k]|k+1,k\rangle$.

At time step $n$, the overall state of the graph and runway will be:

\begin{equation}
\sum_{k=0}^\infty x[k+n]|k+1,k\rangle + \sum_{k=0}^\infty y[n-k-1]|k+1,k\rangle + |G\rangle
\end{equation}

That is, we can describe the coefficients of the edge states on the runway with the ``signal function" $x[n]$ and the ``response function" $y[n]$ which are series of complex numbers, one for each integer $n$.

At the $n$th time step $x[n]$ is the coefficient of $|1,0\rangle$ and $y[n-1]$ is the coefficient of $|0,1\rangle$.  This is so defined such that if $S(z)\equiv r$, where $r$ is constant, then $y[n]=rx[n]$.  When $S(z)$ is constant this holds true for any $x[n]$, but for non-trivial $S(z)$ we have

\begin{equation}\label{ref1}
x[n]=z^n, \forall n \quad \Rightarrow \quad y[n]=S(z)x[n], \forall n
\end{equation}

This is nothing more than a restatement of the definition of $S(z)$, as described in eq. \ref{psi}.  Using $S(z)$ we can derive an expression for the ``impulse response", $h[n]$, which is the response, $y[n]$, produced from the input $x[n] = \delta[n]$, where $\delta[n]=\left\{\begin{array}{ll}1, n=0\\0,n\ne0\end{array}\right.$ is the Kronecker delta function.  From the definitions of $x[n]$, $y[n]$, and the impulse response itself we can say that

\begin{equation}\label{ref2}
h[n]=\langle 0,1|{\bf U}^{n+1}|1,0\rangle
\end{equation}

That is; the impulse is just the state $|1,0\rangle$ at time zero and the impulse response is just the coefficient of $|0,1\rangle$ read at each sequential time step.

We already know that for a simple reflection (such that ${\bf U}|1,0\rangle = r|0,1\rangle$) $y[n]=rx[n]$ and therefore $h[n]=r\delta[n]$.  This can be seen from either of equations \ref{ref1} or \ref{ref2}.  The response of {\it any} signal can be found using the fact that $y[n] = (x*h)[n] = \sum_{k}h[k]x[n-k]$.  So rather than being a single example, the impulse response is the key to finding the response to any input \cite{dspfirst}.

\vspace{5mm}

\begin{thm}
If $h[n]$ is the impulse response (that is, $y[n]=h[n]$ when $x[n]=\delta[n]$), then for any signal function $x[n]$, $y[n] = (x*h)[n]$.
\end{thm}

{\it Proof}

A graph's response to signal functions is a map, $\mathcal{T}$, from the space of sequences of complex numbers to itself.  When we say that $y[n]$ is the response to $x[n]$, we can write this more succinctly as $\mathcal{T}\left(x[n]\right) = y[n]$.  $\mathcal{T}$ inherits linearity from the linearity of ${\bf U}$.

Define $x_k = x[k]$, $\forall k$.  We do this to distinguish between the function $x[n]$ and the value of the function evaluated at $k$.

We can use the Kronecker delta function to break $x[n]$ apart into a sum of simple signals.  Trivially, for any fixed value of $k$, $x_k\delta[n-k] = x_k\delta[k-n] = \left\{\begin{array}{ll}x_k&, n=k\\0&,n\ne k\end{array}\right.$.

Therefore, if we sum over $k$ we reconstruct the full function:

$$x[n] = \sum_k x_k\delta[n-k]$$

From this and the fact that $\mathcal{T}\left(\delta[n]\right) = h[n]$ it follows that

$\begin{array}{ll}
y[n] \\[2mm]
= \mathcal{T}\left(x[n]\right) \\[2mm]
= \mathcal{T}\left(\sum_k x_k\delta[n-k]\right) \\[2mm]
= \sum_k x_k \mathcal{T}\left(\delta[n-k]\right) \\[2mm]
= \sum_k x_k h[n-k] \\[2mm]
= \left(x*h\right)[n]
\end{array}$

$\square$

\vspace{5mm}

\begin{thm}\label{impulse}
The impulse response, $h[n]$, is given by $h[n] = \frac{1}{2\pi i}\oint_{|z|=1} z^{n-1}S(z)\,dz = -\frac{1}{2\pi i}\oint_{|z|=1} z^{n-1}\frac{g(z)}{f(z)}\,dz$
\end{thm}

{\it Proof}

For a single frequency, $x[n]=\lambda^n$, we find that

$\begin{array}{ll}
y[n] = \sum_{k}h[k]x[n-k] \\[2mm]
= \sum_{k}h[k]\lambda^{n-k} \\[2mm]
= \sum_{k}h[k]\lambda^{-k}\lambda^n \\[2mm]
= \left(\sum_{k}h[k]\lambda^{-k}\right)x[n] \\[2mm]
\end{array}$

For any fixed value of $\lambda$ we already have a way of writing this (thm. \ref{important1}).  Therefore

\begin{equation}\label{simpulse}
\sum_{k}h[k]\lambda^{-k} = S(\lambda) = -\frac{g(\lambda)}{f(\lambda)}
\end{equation}

Rewriting $\lambda = e^{i\omega}$:

$\begin{array}{ll}
S(\lambda) = \sum_{k}h[k]\lambda^{-k} \\[2mm]
\Rightarrow \sum_{k}h[k]e^{-i\omega k} = S\left(e^{i\omega}\right) \\[2mm]
\Rightarrow \sum_{k}h[k]e^{i\omega (n-k)} = e^{in\omega}S\left(e^{i\omega}\right)  \\[2mm]
\Rightarrow \int_0^{2\pi} \sum_{k}h[k]e^{i\omega (n-k)}\,d\omega = \int_0^{2\pi} e^{in\omega}S\left(e^{i\omega}\right)  \,d\omega \\[2mm]
\Rightarrow 2\pi h[n] = \int_0^{2\pi} e^{in\omega}S\left(e^{i\omega}\right)  \,d\omega \\[2mm]
\Rightarrow h[n] = \frac{1}{2\pi}\int_0^{2\pi} e^{in\omega} S\left(e^{i\omega}\right) \,d\omega \\[2mm]
\Rightarrow h[n] = \frac{1}{2\pi i}\oint_{|z|=1} z^{n-1}S(z)\,dz & (z=e^{i\omega}, dz = ie^{i\omega}d\omega) \\[2mm]
\Rightarrow h[n] = -\frac{1}{2\pi i}\oint_{|z|=1} z^{n-1}\frac{g(z)}{f(z)}\,dz \\[2mm]
\end{array}$

This is a perfectly nice integral, since the zeros of $f(z)$ are all on the interior of the unit disk.

$\square$

\vspace{5mm}

\begin{thm}\label{impulse2}
For a single input and output, $h[n]= \Omega_0\delta[n-s] + \sum_j \Omega_j\eta_j^n$

where $\Omega_j = \left\{\begin{array}{ll}
-g_0\prod_k \left(\frac{1}{-\eta_k}\right) &, j=0 \\[2mm]
 -g_0\frac{1-|\eta_j|^2}{\eta_j^{s+1}}\prod_{k\ne j}\left(\frac{1-\eta_j\eta_k^*}{\eta_j-\eta_k}\right)  &,  j\ne0
\end{array}\right.$
\end{thm}

{\it Proof}

From the previous theorem the impulse response is:

\begin{equation}
h[n] = -\frac{1}{2\pi i}\oint_{|z|=1} z^{n-1}\frac{g(z)}{f(z)}\,dz = -\frac{g_0}{2\pi i}\oint_{|z|=1} z^{n-s-1}\prod_k\frac{1-z\eta_k^*}{z-\eta_k}\,dz
\end{equation}

\vspace{5mm}

Using residue calculus we can solve this directly

$\begin{array}{ll}
h[n] \\[2mm]
= -\frac{g_0}{2\pi i}\oint_{|z|=1} z^{n-s-1}\prod_k\frac{1-z\eta_k^*}{z-\eta_k}\,dz \\[2mm]
= -g_0\prod_k \left(\frac{1}{-\eta_k}\right)\delta[n-s] - g_0\sum_j (1-|\eta_j|^2)\prod_{k\ne j}\left(\frac{1-\eta_j\eta_k^*}{\eta_j-\eta_k}\right) \eta_j^{n-s-1} \\[2mm]

= -g_0\prod_k \left(\frac{1}{-\eta_k}\right)\delta[n-s] + \sum_j \left[-g_0\frac{1-|\eta_j|^2}{\eta_j^{s+1}}\prod_{k\ne j}\left(\frac{1-\eta_j\eta_k^*}{\eta_j-\eta_k}\right)\right] \eta_j^n
\end{array}$

$\square$

\vspace{5mm}

So the impulse response is a set of exponentially decaying signals corresponding to the zeros of $f(z)$.

\pagebreak

\section{Quantum Sounding}

Since there is a closed form for the response to any signal that is a function only of the eigenvalues of ${\bf U}_0$, we can (at most) find the spectrum of a graph's eigenvalues.  We are now equipped to ask the question ``What can be learned about a graph attached to a runway by means of a signal on that runway?" and even ask the more practical question ``How difficult is it to do so?".

The challenge we face is that the closer $\eta$ is the unit circle, the more difficult it is to detect.  From eq. \ref{compliment} we know that (for a single runway) $S(z) = -\frac{g(z)}{f(z)} = -\frac{g_0}{z^s} \prod_k \frac{1 - z\eta_k^*}{z-\eta_k}$.  When $|z|=1$ we can easily verify that $\left| S(z) \right| = 1$.  More specifically, for each $k$, $\left| \frac{1 - z\eta_k^*}{z-\eta_k} \right| = 1$.  We'll look at each of these individually and since each has modulus 1, we can concern ourselves entirely with their phase.

Here we assume that $0\ll|\eta_k|<1$.  When $z\not\approx \eta_k$ we can see that $\frac{1 - z\eta_k^*}{z-\eta_k} = -\eta_k^* \left( \frac{z - \frac{1}{\eta_k^*}}{z-\eta_k} \right) \approx -\eta_k^*$.

Importantly, $\frac{1 - z\eta_k^*}{z-\eta_k}$ is approximately constant for most values of $z$ on the unit circle.  This is because $\frac{1}{\eta_k^*} = \frac{\eta_k}{\left|\eta_k\right|^2}\approx\eta_k$, and therefore $\left( \frac{z - \frac{1}{\eta_k^*}}{z-\eta_k} \right)\approx 1$.

$\frac{1 - z\eta_k^*}{z-\eta_k}$ has one pole inside of the unit circle and one zero outside of it.  As a result, when we apply the argument principle we find that if $z$ runs around the unit circle in a positively oriented loop, then $\Delta\arg\left( \frac{1 - z\eta_k^*}{z-\eta_k} \right) = -2\pi$.

\vspace{5mm}

It follows then that a zero of $f(z)$ has very little impact on $S(z)$ except when $z$ is within a small neighborhood of that zero and within that small neighborhood the phase suddenly jumps by $-2\pi$.  We'll now make this a little more rigorous to find the extent of this neighborhood.

\vspace{5mm}

\begin{thm}\label{window}
If $\eta=(1-\delta)e^{i\tau}$ is a root of $f(z)$ and $x[n]=e^{in\theta}$, we find that $\eta$ can only be detected when $|\theta-\tau|=O(\delta)$.  Moreover, the phase of the reflection coefficient decreases by $2\pi$ in this neighborhood of $\eta$.
\end{thm}

{\it Proof}

Without loss of generality, we can assume that $\eta=1-\delta$.  We define $e^{i\phi} = \frac{1 - z\eta_k^*}{z-\eta_k} = \frac{1 - e^{i\theta}(1-\delta)}{e^{i\theta}-(1-\delta)}$ and quickly find that

$$e^{i\phi} = \left[-1+\frac{\delta^2(1+\cos{(\theta)})}{2(1-\delta)(1-\cos{(\theta)})+\delta^2}\right] + i\frac{(-2\delta+\delta^2)\sin{(\theta)}}{2(1-\delta)(1-\cos{(\theta)})+\delta^2}$$

Clearly, for $\theta\ne0$, $\lim_{\delta\to0}e^{i\phi}=-1$.  Define $c=\frac{\theta}{\delta}=O(1)$.  The imaginary part of this last equation for small values of $\delta$ and $\theta$ is

$$\sin{(\phi)} = \frac{(-2\delta+\delta^2)\sin{(c\delta)}}{2(1-\delta)(1-\cos{(c\delta)})+\delta^2} = \frac{-2c\delta^2+O(\delta^3)}{c^2\delta^2+\delta^2 +O(\delta^3)} = -\frac{2c}{c^2+1} + O(\delta)$$

So the window for which $|\phi|<\frac{\pi}{2}$ is approximately $-\delta<\theta<\delta$ and the window for which $|\phi-\pi|>d$ (for which $e^{i\phi}$ is different from -1) is approximately $-\frac{2}{d}\delta<\theta<\frac{2}{d}\delta$.

This is the statement of the theorem.

$\square$

\vspace{5mm}

In appendix \ref{valve} there is an example of how a graph structure can be used to detect the phase of a reflection coefficient.

\vspace{5mm}

From theorem \ref{polynomial} we know that $b(z)f(z) = \left| {\bf U}_0 -z{\bf I} \right|$, which means that the degree of the polynomial $f(z)$ is less than or equal to the number of edge states in $G$.  $S(z) = -\frac{g(z)}{f(z)}$, as established in theorem \ref{important1}.  Since the zeros of $f(z)$ are all within the unit circle and the zeros of $g(z)$ are all outside, we can apply the argument principle to $S(z)$ around the unit circle to conclude that

$$\left|\Delta\arg\left(S(z)\right)\right|\le 2\pi|G|$$

The equality is achieved if there are no bound states.  So, with very little effort, we have an algorithm that provides a lower bound for the dimension of a graph's Hilbert space (the number of edge states).

\pagebreak

\section{Scattering: Multiple Inputs and Outputs}

Define ${\bf U}_0$ to be the time step operator of a {\it finite} graph, $G$, with some states prepared as either loose inputs or outputs.  An input state has no pre-image and an output state has no image.  That is: ${\bf U}_0|out\rangle = 0$ and ${\bf U}_0^{-1}|in\rangle = \emptyset$.

If we wish to ``splice" a runway onto the graph we first choose one input state, $|in_j\rangle$, and one output state, $|out_k\rangle$, and define ${\bf U}^{(jk)}$ as ${\bf U}^{(jk)}|1,0\rangle = |in_j\rangle$ and ${\bf U}^{(jk)}|out_k\rangle = |0,1\rangle$.  The behavior on the runway is the same regardless of which states on $G$ it connects with, so $\forall j,k$ and $n>0$, ${\bf U}^{(jk)} |n-1,n\rangle = |n,n+1\rangle$ and ${\bf U}^{(jk)} |n+1,n\rangle = |n,n-1\rangle$.

\vspace{5mm}

A $\lambda$-eigenstate of ${\bf U}^{(jk)}$ necessarily takes the form:

\begin{equation}
|\Psi\rangle = \sum_{n=0}^\infty \Big[ \lambda^{n+1} |n+1,n\rangle + S_{jk}(\lambda) \lambda^{-n} |n,n+1\rangle \Big] + |in_j\rangle + \lambda S_{jk}(\lambda) |out_k\rangle + \cdots
\end{equation}

As before we introduce another operator, ${\bf U}^{(jk)}_\alpha \equiv {\bf U}_0 + \alpha |in_j\rangle\langle out_k|$, that reflects back into the graph rather than communicating with the runway.  For both of these operators ${\bf U}^{(jk)}_\alpha |out_i\rangle = {\bf U}^{(jk)} |out_i\rangle = 0$, $\forall i\ne k$.

A $\lambda$-eigenstate of ${\bf U}^{(jk)}_\alpha$ necessarily takes the form:

\begin{equation}
|\Psi_\alpha\rangle = |in_j\rangle + \frac{\lambda}{\alpha} |out_k\rangle + \cdots
\end{equation}

If the $\lambda$-eigenstates are identical on $G$, then clearly $\frac{\lambda}{\alpha} = \lambda S_{jk}(\lambda)$ and therefore $S_{jk}(\lambda) = \frac{1}{\alpha}$.

\vspace{5mm}

\begin{thm}
$S_{jk}(\lambda) = -\frac{g_{jk}(\lambda)}{f(\lambda)}$, where $C_{jk}(z) = \left| {\bf U}^{(jk)}_\alpha - z{\bf I} \right| = f(z) + \alpha g_{jk}(z)$ is the characteristic polynomial of ${\bf U}^{(jk)}_\alpha$.
\end{thm}

{\it Proof}  If $\lambda$ is an eigenvalue of ${\bf U}^{(jk)}_\alpha$, then $0 = f(\lambda) + \alpha g_{jk}(\lambda)$.  By matching the coefficient of the $|out_k\rangle$ state in $|\Psi_\alpha\rangle$ and $|\Psi\rangle$ we find that $S_{jk}(\lambda) = \frac{1}{\alpha} = -\frac{g_{jk}(\lambda)}{f(\lambda)}$.

$\square$

\vspace{5mm}

This is essentially the same as thm. \ref{important1} with one unimportant difference.  In this case ${\bf U}^{(jk)}_\alpha$ is no longer unitary since ${\bf U}^{(jk)}_\alpha|out_\ell\rangle=0$ for $\ell\ne k$.  As a result we can no longer say that $|\alpha|=1$, however this has no impact on the proof.  In fact, it is to be expected that for multiple runways $|\alpha|\ge1$.  From thm. \ref{ssum} we see that for $|z|=1$, $\sum_k |S_{jk}(z)|^2=1$ so we can conclude that $\left|S_{jk}(z)\right|\le1$ and $|\alpha|=\frac{1}{\left|S_{jk}(z)\right|}\ge1$.

\vspace{5mm}

For any operator ${\bf M}$ we call $\left({\bf M} - z{\bf I}\right)^{-1}$ the ``resolvent" of ${\bf M}$.  The resolvent has many fascinating properties \cite{kato}, and here we introduce one more.

\vspace{5mm}

\begin{thm}[The Resolvent Theorem]\label{important2}
$S_{jk}(z) = -\langle out_k| \left({\bf U}_0 - z{\bf I}\right)^{-1} |in_j\rangle$ or stated differently ${\bf S}^{T}(z) = -\left({\bf U}_0 - z{\bf I}\right)^{-1}$.
\end{thm}

{\it Proof} By removing the minor of the $\alpha$ element of ${\bf U}^{(jk)}_\alpha$ we have that $\left|{\bf U}^{(jk)}_\alpha-z{\bf I}\right| = \left|{\bf U}_0 - z{\bf I}\right| + \alpha\left|{\bf U}_\alpha^{(jk)} - z{\bf I}\right|_{<j,k>} = \left|{\bf U}_0 - z{\bf I}\right| + \alpha\left|{\bf U}_0-z{\bf I}\right|_{<j,k>}$, where the sub-index indicates the cofactor of the $\alpha$ element (the $|in_j\rangle\langle out_k|$ element) in ${\bf U}_\alpha^{(jk)}$.

So, $\left\{\begin{array}{ll}
f(z) = \left|{\bf U}_0 - z{\bf I}\right|  \\
g_{jk}(z) = \left|{\bf U}_0 - z{\bf I}\right|_{<j,k>} \end{array}\right.$ 

$\left|{\bf U}^{(jk)}_\alpha-z{\bf I}\right|_{<j,k>} = \left|{\bf U}_0-z{\bf I}\right|_{<j,k>}$, since this cofactor is not a function of $\alpha$ (that row and column is removed).  It is a known property of cofactors that if ${\bf B}$ is the cofactor matrix of ${\bf A}$ (that is; every element of ${\bf B}$ is the corresponding cofactor of ${\bf A}$), then $\left|{\bf A}\right|{\bf A}^{-1} = {\bf B}^{T}$.  It follows that:

\begin{equation}\label{transpose}
S_{jk}(z) = -\frac{g_{jk}(\lambda)}{f(\lambda)} = -\frac{\left|{\bf U}_0-z{\bf I}\right| \left({\bf U}_0-z{\bf I}\right)^{-1}_{kj}}{\left|{\bf U}_0-z{\bf I}\right|} = -\left({\bf U}_0-z{\bf I}\right)^{-1}_{kj}\end{equation}

This applies only to those edges that are ``prepared" to be inputs and outputs as described at the beginning of this section.   That is, $\alpha$ needs to be the only element appearing in both its row and column.  Replacing an arbitrary element of ${\bf U}_0$ with $\alpha$ destroys the unitarity of the time step operator, ${\bf U}$.

\vspace{5mm}

Equation \ref{transpose} says that when you want $S_{jk}(\lambda)$, the scattering coefficient between $|in_j\rangle$ and $|out_k\rangle$, you can find it in the $|out_k\rangle\langle in_j|$ element of $\left({\bf U}_0-z{\bf I}\right)^{-1}$.  In other words:

\begin{equation}
S_{jk}(z) = -\langle out_k| \left({\bf U}_0 - z{\bf I}\right)^{-1} |in_j\rangle
\end{equation}

$\square$

\vspace{5mm}

The above theorem applies to the single runway case as well.

\subsection{Particulars for Multiple Runways}

Once again, define ${\bf U}_\alpha^{(jk)} \equiv {\bf U}_0 + \alpha |in_j\rangle\langle out_k|$ and $C_{jk}(z) = \left|{\bf U}_\alpha^{(jk)} - z{\bf I}\right| = b(z)\left(f(z) + \alpha g_{jk}(z)\right)$.

The zeros of $g_{jk}(z)$ are {\it not} fixed by the zeros of $f(z)$, the way there are in the single runway case, and are not necessarily outside of the unit circle.  We can see an example of this in appendix \ref{square}.

As in thm. \ref{polynomial}, those eigenvalues with modulus 1 correspond to bound eigenstates, and are factored out as $b(z)$.  However, in the general case we find that $g_{jk}(z)$ and $f(z)$ may have zeros in common.  If $|\psi\rangle$ is an eigenstate such that $\langle out_k|\psi\rangle = \langle in_j|\psi\rangle = 0$, then it is independent of $\alpha$, however it may not necessarily be a bound state since there are other runways that it can be ``leaking" out of.  If an eigenvalue is independent of $\alpha$ then it must be a common factor of $f(z)$ and $g_{jk}(z)$ and if the associated eigenstate has a component on any of the $|out\rangle$ states, then its eigenvalue must be less than 1.  In fact, in a derivation nearly identical to that found in theorem \ref{polynomial} we find that the eigenvalue, $\eta$, of a eigenstate of ${\bf U}_0$, $|\psi\rangle$, satisfies $|\eta|^2 = 1 - \sum_{k=1}^M \left|\langle out_k|\psi\rangle\right|^2 < 1$

These are the zeros of $f(z)$.  If $\langle out_k|\psi\rangle=0$, then $\eta$ is also a zero of $g_{jk}(z)$ and $|\psi\rangle$ is an eigenstate of ${\bf U}_\alpha^{(jk)}$, $\forall j$.

\vspace{5mm}

\begin{thm}\label{ssum}
If $|z|=1$, then $\sum_{k}\left|S_{jk}(z)\right|^2 = 1$.
\end{thm}

{\it Proof} The $\lambda$-eigenstate for a graph $G$ attached to $M$ runways with a signal coming in from the $j$th runway takes the form

$$|\Psi\rangle=\sum_{n=0}^\infty \lambda^{n+1} |n+1,n\rangle_j + \sum_{k=1}^M\left[S_{jk}(\lambda)\sum_{n=0}^\infty \lambda^{-n} |n,n+1\rangle_k\right] + |G\rangle + |in_j\rangle + \sum_{k=1}^M \lambda S_{jk}(\lambda)|out_k\rangle$$

The subscript on the runway states indicates to which runway they correspond.  This is an eigenstate, so ${\bf U}|\Psi\rangle = \lambda|\Psi\rangle$ and it follows that ${\bf U}\left(|G\rangle + |in_j\rangle\right) = \lambda\left(|G\rangle + \sum_{k=1}^M \lambda S_{jk}(\lambda)|out_k\rangle\right)$.  Being unitary ${\bf U}$ is an isometry and therefore

$\begin{array}{ll}
\left||G\rangle + |in_j\rangle\right| =\left|\lambda|G\rangle + \lambda^2\sum_{k=1}^M S_{jk}(\lambda)|out_k\rangle\right| \\[2mm]
\Rightarrow 1 + \langle G|G\rangle = |\lambda|^2 \langle G|G\rangle + |\lambda|^4 \sum_{k=1}^M \left|S_{jk}(\lambda)\right|^2
\end{array}$

Clearly if $|\lambda|=1$, then the statement of the theorem follows immediately.

\vspace{5mm}

It may seem worrisome that $\lambda$ isn't assumed to be modulus 1 despite being the eigenvalue of a unitary operator, but keep in mind that $|\Psi\rangle$ isn't normalizable and ${\bf U}$ is operating on an infinite dimensional Hilbert space.  However, on any {\it finite} state we can still make use of the fact that unitary operations are isometries.

$\square$

\subsection{Arbitrary Signals for Multiple Runways}

As before, define the input $x[n]$ (output $y[n]$) as the amplitude on the state $|1,0\rangle$ (state $|0,1\rangle$) at time step $n$ (time step $n+1)$.  The input can be encoded onto the runway in an initial state of the form $\sum_{n=0}^\infty x[n] |n+1,n\rangle$.  If $x[n]=\lambda^n$ for all $n$, then $y[n] = S_{jk}(\lambda)x[n]$.

Applying exactly the same proof used in the single input/output case (thm. \ref{impulse}) we find that 

\begin{equation}
h_{jk}[n] = \frac{1}{2\pi i}\oint_{|z|=1} z^{n-1}S_{jk}(z)\,dz =  -\frac{1}{2\pi i}\oint_{|z|=1} z^{n-1}\frac{g_{jk}(z)}{f(z)}\,dz
\end{equation}

where $h_{jk}[n]$ is the response produced by the $k$th output to an impulse received from the $j$th input.

\vspace{5mm}

Like the single runway case we find that $S_{jk}(z) = \sum_{n=0}^{\infty} h_{jk}[n]z^{-n}$ (see eq. \ref{simpulse}).  This gives us a second proof and a little insight into theorem \ref{important2}.

In general, the impulse response is $h_{jk}[n] = \langle 0,1|\left({\bf U}^{(jk)}\right)^{n+1}|1,0\rangle$ (see eq. \ref{ref2}), where ${\bf U}^{(jk)}|1,0\rangle = |in_j\rangle$ and ${\bf U}^{(jk)}|out_k\rangle = |0,1\rangle$.  The way we have defined the graph (such that it includes ``in" and ``out" states) implies that $h_{jk}[0]=h_{jk}[1]=0$, and therefore $S_{jk}(z) = \sum_{n=2}^{\infty} h_{jk}[n]z^{-n}$.  In what follows we'll keep $h_{jk}[1]$ in the sum; this changes nothing but makes the derivation a little smoother.

\vspace{5mm}

$\begin{array}{ll}
S_{jk}(z) \\[2mm]
= \sum_{n=1}^{\infty} h_{jk}[n]z^{-n} \\[2mm]
= \sum_{n=1}^{\infty} z^{-n}\langle 0,1|\left({\bf U}^{(jk)}\right)^{n+1}|1,0\rangle \\[2mm]
= \sum_{n=1}^{\infty} z^{-n}\langle out_k|\left({\bf U}^{(jk)}\right)^{n-1}|in_j\rangle \\[2mm]
= \sum_{n=1}^{\infty} z^{-n}\langle out_k|{\bf U}_0^{n-1}|in_j\rangle \\[2mm]
= \sum_{n=0}^{\infty} z^{-n-1}\langle out_k| {\bf U}_0^n |in_j\rangle \\[2mm]
= \langle out_k| \left[ \sum_{n=0}^{\infty} z^{-n-1} {\bf U}_0^n \right] |in_j\rangle \\[2mm]
= \langle out_k| \left[ \frac{1}{z} \sum_{n=0}^{\infty} \left( \frac{1}{z} {\bf U}_0 \right)^n \right] |in_j\rangle \\[2mm]
= \langle out_k| \left[ \frac{1}{z} \left( {\bf I} - \frac{1}{z} {\bf U}_0 \right)^{-1} \right] |in_j\rangle \\[2mm]
= \langle out_k| \left[ - \left( {\bf U}_0 - z{\bf I} \right)^{-1} \right] |in_j\rangle \\[2mm]
\end{array}$

\vspace{5mm}

So, we can either think of theorem \ref{important2} as being a result of the nature of the characteristic polynomial of ${\bf U}_\alpha$ or as a symptom of the fact that the power series of the negative resolvent, $-\left( {\bf U}_0 - z{\bf I} \right)^{-1}$, is a sum of impulse responses multiplied by $z^{-n}$ which is equal to the frequency response, $S(z)$.

\pagebreak

\section*{Acknowledgements}

This research was supported by a grant from the John Templeton Foundation and would not have been possible without many enlightening conversations with Professor Mark Hillery.

\pagebreak

\pagebreak

\appendix

\section{Appendix}

The goal of this appendix is to present a series of elucidating examples and to demonstrate and verify the theorems presented in the paper.

The bolo graph example is a single runway graph that's simple enough for the theorems described in this paper to be verified directly through brute force calculation.

The star graph and complete graph examples reiterate some of the basic theorems from section 2 as well as demonstrate how the structure of a graph can be discerned using signals.

The simple valve and square junction examples are multiple-runway examples.  The simple valve has two runways and we see how we can either attain total reflection or transmission by changing the graph or the signal.  The square junction has four runways and is included to show how easily the techniques can be generalized. 

The pruned tree example shows how, in practice,  the scattering coefficients of a subgraph can be used to replace it in a larger graph.

\subsection{Example: The Bolo Graph}

The bolo graph is among the smaller non-trivial graphs.  It has one bound eigenstate and is named for its resemblance to a bolo tie.

\begin{figure}[h!]
\centering
\includegraphics[width=2.7in]{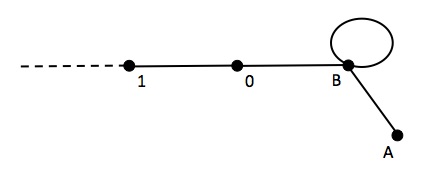}
\caption{The bolo graph attached to an infinite runway.  Here $|in\rangle = |0,B\rangle$ and $|out\rangle = |B,0\rangle$.  That is, ${\bf U}|1,0\rangle=|0,B\rangle$ and ${\bf U}|B,0\rangle=|0,1\rangle$ whereas ${\bf U}_\alpha|B,0\rangle=\alpha|0,B\rangle$.}
\end{figure}

The states on the Bolo graph are

$\begin{array}{ll}
|\psi_1\rangle = |0,B \rangle = |in\rangle \\
|\psi_2\rangle = |A,B \rangle\\
|\psi_3\rangle = |B,B \rangle\\
|\psi_4\rangle = |B,A \rangle\\
|\psi_5\rangle = |B,0 \rangle = |out\rangle
\end{array}$

and ${\bf U}_\alpha$ is defined to act on these states as

\vspace{5mm}

${\bf U}_\alpha=\left(\begin{array}{ccccccc}
0&0&0&0&\alpha\\
0&0&0&-1&0\\
\frac{2}{3}&\frac{2}{3}&-\frac{1}{3}&0&0\\
\frac{2}{3}&-\frac{1}{3}&\frac{2}{3}&0&0\\
-\frac{1}{3}&\frac{2}{3}&\frac{2}{3}&0&0
\end{array}\right)$

The characteristic polynomial, written using the form described in theorem \ref{polynomial}, is

$$\left|{\bf U}_\alpha - z{\bf I}\right| = -\underbrace{(z+1)}_{b(z)}\left[\underbrace{z^2 \left(z^2-\frac{2}{3}z+\frac{1}{3}\right)}_{f(z)} + \alpha \underbrace{\left(\frac{1}{3}z^2-\frac{2}{3}z+1\right)}_{g(z)}\right]$$

In this form it's clear that $f(z)$ and $g(z)$ are reciprocal polynomials.  We quickly find that $\eta = \left\{\frac{1+i\sqrt{2}}{3}, \frac{1-i\sqrt{2}}{3}\right\}$, $s=2$, and $g_0 = 1$.

\vspace{5mm}

This has five roots and when $|\alpha|=1$ all five have modulus 1, as they should since ${\bf U}_\alpha$ is unitary when $|\alpha|=1$.  $|\eta| = \left|\frac{1\pm i\sqrt{2}}{3}\right| = \frac{1}{\sqrt{3}}$, so the roots of $f(z)$ all fall inside the unit circle (thm. \ref{polynomial}).

Graphing the magnitude of the relevant part of the polynomial (ignoring the ``$z+1$") demonstrates the relationship between $\alpha$ and the zeros as described by theorems \ref{distinct}, \ref{permute}, and \ref{unique} (see fig. 4).

\begin{figure}[h!]
\centering
\includegraphics[width=3.5in]{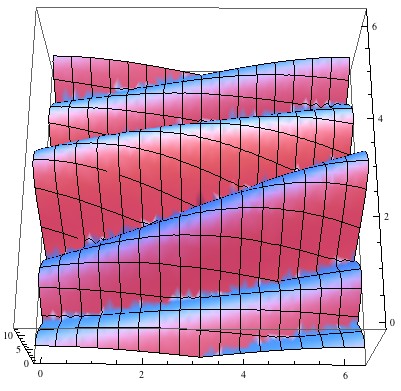}
\caption{$\left|f(z)+\alpha g(z)\right|$ for the bolo graph where $\alpha=e^{ix}$ and $z=e^{iy}$.  The zeros are clearly visible as troughs.  Notice also that for a given zero there is one corresponding value of $\alpha$.}
\end{figure}

\vspace{5mm}

According to theorem \ref{important1} the effective reflection coefficient off of vertex 0 is

\begin{equation}\label{boloscatter}
S(z) = -\frac{g(z)}{f(z)} = -\frac{z^2-2z+3}{z^2(3z^2-2z+1)}
\end{equation}

\vspace{5mm}

We can find $S(\lambda)$ by brute force {\it one eigenvalue at a time} by calculating the appropriate eigenstate.  For $\lambda=i$, we find that the eigenstate is:

\begin{equation}
|\Psi\rangle=\sum_{j=0}^\infty i^{j+1}|j+1,j\rangle+\sum_{j=0}^\infty i^{1-j}|j,j+1\rangle+|0,B\rangle-|B,0\rangle-\frac{1+i}{2}|B,A\rangle+\frac{1-i}{2}|A,B\rangle-i|B,B\rangle
\end{equation}

Comparing carefully with the definition of $S(\lambda)$ described in section 2, we find that $S(i)=i$.  This lines up exactly with what we should expect, since (from eq. \ref{boloscatter}) $S(i) = -\frac{i^2-2i+3}{i^2(3i^2-2i+1)} = i$.

\vspace{5mm}

Often we would prefer not to have an extra edge on the runway.  In this case that edge is between $B$ and $0$.  In order to find the effective reflection coefficient of vertex $B$ instead of vertex $0$ we need to delay the signal by two time steps.  This is accomplished easily through a multiplication by $z^2$.  The effective reflection coefficient of vertex $B$ is

\begin{equation}
R(z) = -z^2\frac{g(z)}{f(z)} = -\frac{z^2-2z+3}{3z^2-2z+1}
\end{equation}

\begin{figure}[h!]
\centering
\includegraphics[width=3.5in]{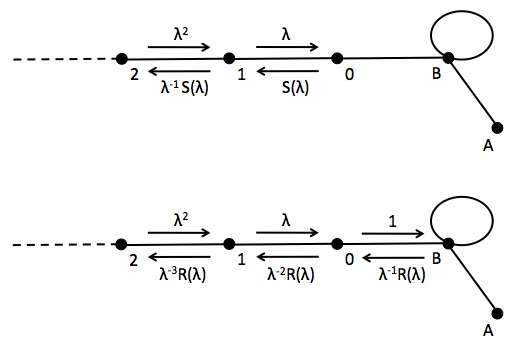}
\caption{$S(\lambda)$ is defined normally, as the effective reflection coefficient of vertex $0$.  $R(\lambda)$ is here defined as the effective reflection coefficient of vertex $B$.  From the diagram it's straightforward to see that $R(z) = z^2S(z)$.}
\label{Rdef}
\end{figure}

\vspace{5mm}

Although the bolo graph only has one input and output state, we can still find the scattering/reflection coefficient using the resolvent, as described in theorem \ref{important2}.  Using the same basis states as before, we find that

${\bf U}_0=\left(\begin{array}{ccccccc}
0&0&0&0&\color{red}{0}\\
0&0&0&-1&0\\
\frac{2}{3}&\frac{2}{3}&-\frac{1}{3}&0&0\\
\frac{2}{3}&-\frac{1}{3}&\frac{2}{3}&0&0\\
-\frac{1}{3}&\frac{2}{3}&\frac{2}{3}&0&0
\end{array}\right)$

Here the $|in\rangle\langle out| = |\psi_1\rangle\langle \psi_5|$ element has been marked in red.  We find that the negative resolvent is

$-\left({\bf U}_0 - z{\bf I}\right)^{-1} = \left(\begin{array}{ccccc}
\frac{1}{z} & 0 & 0 & 0 & 0 \\[2mm]
-\frac{2}{z(1 - 2 z + 3 z^2)} & \frac{z (1 + 3 z)}{1 - z + z^2 + 3 z^3} & -\frac{2}{1 - z + z^2 + 3 z^3} & -\frac{1 + 3 z}{1 - z + z^2 + 3 z^3} & 0 \\[2mm]
\frac{2 (-1 + z)}{z (1 - 2 z + 3 z^2)} & \frac{2 z}{1 - z + z^2 + 3 z^3} & \frac{-1 + 3 z^2}{1 - z + z^2 + 3 z^3} & -\frac{2}{1 - z + z^2 + 3 z^3} & 0 \\[2mm]
\frac{2}{1 - 2 z + 3 z^2} & \frac{1 - z}{1 - z + z^2 + 3 z^3} & \frac{2 z}{1 - z + z^2 + 3 z^3} & \frac{z (1 + 3 z)}{1 - z + z^2 + 3 z^3} & 0 \\[2mm]
\color{red}{-\frac{3 - 2 z + z^2}{z^2(1 - 2 z + 3 z^2)}} & \frac{2}{1 - 2 z + 3 z^2} & \frac{2 (-1 + z)}{z (1 - 2 z + 3 z^2)} & -\frac{2}{z(1 - 2 z + 3 z^2)} & \frac{1}{z} \\[2mm]
\end{array}\right)$

The scattering coefficient is the $|out\rangle\langle in| = |\psi_5\rangle\langle \psi_1|$ element and is again marked in red.  As expected (thm. \ref{important2}) this is precisely the result derived earlier in this example.

\vspace{5mm}

Finally, we look at the impulse response of the bolo graph.  The first few terms of the impulse response can be found by direct calculation:

$\begin{array}{c|c|c|c}
 n & {\bf U}^n|1,0\rangle & x[n]& y[n] \\\hline
0 & |1,0\rangle & 1 & 0 \\[2mm]
1 & |0,B\rangle & 0 & 0 \\[2mm]
2 & \frac{2}{3}|B,B\rangle + \frac{2}{3}|B,A\rangle - \frac{1}{3}|B,0\rangle & 0 & -\frac{1}{3} \\[2mm]
3 & - \frac{2}{3} |A,B\rangle - \frac{2}{9}|B,B\rangle + \frac{4}{9}|B,A\rangle + \frac{4}{9}|B,0\rangle - \frac{1}{3}|0,1\rangle  & 0 & \frac{4}{9} \\[2mm]
4 & - \frac{4}{9} |A,B\rangle - \frac{10}{27}|B,B\rangle + \frac{2}{27}|B,A\rangle - \frac{16}{27}|B,0\rangle + \frac{4}{9}|0,1\rangle - \frac{1}{3}|1,2\rangle & 0 & - \frac{16}{27} \\[2mm]

5 & - \frac{2}{27} |A,B\rangle - \frac{14}{81}|B,B\rangle - \frac{8}{81}|B,A\rangle - \frac{44}{81}|B,0\rangle - \frac{16}{27}|0,1\rangle + \frac{4}{9}|1,2\rangle - \frac{1}{3}|2,3\rangle & 0 & - \frac{44}{81} \\[2mm]

\end{array}$

\vspace{5mm}

Keep in mind here that $y[n]=S(\lambda)x[n]$ when $x[n]=\lambda^n$.  As a result of this definition $y[n]$ ``anticipates" when you write out a list like this: during the $n$th time step $x[n]$ is the coefficient of $|1,0\rangle$ and $y[n-1]$ is the coefficient of $|0,1\rangle$.

\vspace{5mm}

Rather than an infinite brute force calculation, we can apply theorem \ref{impulse2}.  Plugging in $\eta_1 = \frac{1+i\sqrt{2}}{3}$, $\eta_2 = \frac{1-i\sqrt{2}}{3}$, $s=2$, and $g_0 = 1$ we find:

$\begin{array}{ll}
\Omega_0 = -\left(\frac{-3}{1+i\sqrt{2}}\right)\left(\frac{-3}{1-i\sqrt{2}}\right) = -3 \\[2mm]
\Omega_1 = -\frac{1-\left|\frac{1+i\sqrt{2}}{3}\right|^2}{\left(\frac{1+i\sqrt{2}}{3}\right)^3}\left(\frac{1-\left(\frac{1+i\sqrt{2}}{3}\right)\left(\frac{1+i\sqrt{2}}{3}\right)}{\left(\frac{1+i\sqrt{2}}{3}\right)-\left(\frac{1-i\sqrt{2}}{3}\right)}\right) = -i3\sqrt{2}\\[2mm]
 \Omega_2 = -\frac{1-\left|\frac{1-i\sqrt{2}}{3}\right|^2}{\left(\frac{1-i\sqrt{2}}{3}\right)^3}\left(\frac{1-\left(\frac{1-i\sqrt{2}}{3}\right)\left(\frac{1-i\sqrt{2}}{3}\right)}{\left(\frac{1-i\sqrt{2}}{3}\right)-\left(\frac{1+i\sqrt{2}}{3}\right)}\right) = i3\sqrt{2}
 \end{array}$

\vspace{5mm}

So for $n\ge2$,

$h[n] = -3 \delta[n-2] - i3\sqrt{2}\left(\frac{1+i\sqrt{2}}{3}\right)^n + i3\sqrt{2}\left(\frac{1-i\sqrt{2}}{3}\right)^n$

\subsection{Example: A Star Graph with Differently Marked Edges}

The only information we can ever hope to gain from a signal are the zeros of $f(z)$ and the value of $g_0$.  Every other quantity described so far can be determined from these.

However, a graph with $d$ edge states can have any of a huge number of arrangements, and each vertex in that graph can have any appropriate unitary operator.  It is completely unreasonable to hope for a way of determining the structure of a completely arbitrary graph by studying its responses to signals.  That is to say, using the techniques described in this paper we cannot hope to distinguish between isospectral graphs, and therefore cannot expect to uniquely determine the structure of an unknown graph strictly from its response to signals.

However, if we restrict the graph to being one of a restricted family of graphs, then we can expect to get some information about the structure.  First we investigate a star graph.

\vspace{5mm}

Assume that we know that there are $N+1$ edges radiating from a central vertex $c$ which is diffusive ($r=-1+\frac{2}{N+1}$ for every edge and $t=\frac{2}{N+1}$ between every pair of edges).

The terminating vertex of some of $M$ of these flip the phase (multiply by -1) when they reflect, $N-M$ of them leave the phase unchanged (multiply by 1), and the last edge connects to the runway.  The question is: can we determine the number of each type of edge?

Define the states

$\begin{array}{ll}
|\psi_1\rangle = |0,c\rangle = |in\rangle \\
|\psi_2\rangle = |c,0\rangle = |out\rangle \\
|\psi_3\rangle = \frac{1}{\sqrt{M}}\sum_{j=1}^M |c,a_j\rangle \\
|\psi_4\rangle = \frac{1}{\sqrt{M}}\sum_{j=1}^M |a_j,c\rangle \\
|\psi_5\rangle = \frac{1}{\sqrt{N-M}}\sum_{j=1}^{N-M} |c,b_j\rangle \\
|\psi_6\rangle = \frac{1}{\sqrt{N-M}}\sum_{j=1}^{N-M} |b_j,c\rangle \\
\end{array}$

\vspace{5mm}

Here the $a_j$ vertices reflect with $-1$ and the $b_j$ vertices reflect with $1$.  The time step operator can be written

${\bf U}_\alpha = \left(\begin{array}{cccccc}
0 & -1+ \frac{2}{N+1} & 0 & 2\frac{\sqrt{M}}{N+1} & 0 & 2\frac{\sqrt{N-M}}{N+1} \\
\alpha & 0 & 0 & 0 & 0 & 0 \\
0 & 2\frac{\sqrt{M}}{N+1} & 0 & -1+ \frac{2M}{N+1} & 0 & 2\frac{\sqrt{M(N-M)}}{N+1} \\
0 & 0 & -1 & 0 & 0 & 0 \\
0 & 2\frac{\sqrt{N-M}}{N+1} & 0 & 2\frac{\sqrt{M(N-M)}}{N+1} & 0 & -1+ \frac{2(N-M)}{N+1} \\
0 & 0 & 0 & 0 & 1 & 0
\end{array}\right)$

\vspace{5mm}

The characteristic polynomial is

$$C(z,\alpha) = z^6 + \left( \alpha - \frac{2\alpha}{N+1} + \frac{4M}{N+1} - \frac{2 N}{N+1} \right) z^4 + \left( -1 + \frac{4M\alpha}{N+1} + \frac{2 N}{N+1} - \frac{2N\alpha}{N+1} \right)z^2 + \alpha$$

$$= \underbrace{z^2\left( z^4 + \left( \frac{4M-2 N}{N+1} \right) z^2 + \left(\frac{N-1}{1+ N}\right)\right)}_{f(z)} + \alpha\underbrace{\left( \left(\frac{N-1}{ N+1}\right) z^4 + \left( \frac{4M - 2N}{N+1} \right)z^2 + 1\right)}_{g(z)}$$

\vspace{5mm}

The effective reflection coefficient of vertex $c$ is

$$R(z) = z^2 S(z) = -\frac{(N-1) z^4 + \left(4M - 2N\right)z^2 + (N+1)}{(N+1)z^4 + \left(4M-2 N\right) z^2 + \left(N-1\right)}$$

\vspace{5mm}

Where the extra $z^2$ is a delay necessary when the 2 extra states, $|0,c\rangle$ and $|c,0\rangle$, are included in the runway.  It is not immediately obvious how this function behaves.  However, we can hope to understand it better by looking at the zeros of $f(z)$ and applying theorem \ref{window}.  The four non-zero roots of $f(z)$ are

\begin{equation}
\eta = \mathring{\pm} \sqrt{ 1 - \frac{2M + 1}{N+1} \pm \frac{\sqrt{1 - 4M(N-M)}}{N+1}}
\end{equation}

where $\mathring{\pm}$ and $\pm$ are independent.

\vspace{5mm}

Define $L\equiv\frac{M}{N}$, then 

$\begin{array}{ll}
\eta = \mathring{\pm} \sqrt{(1 -2L)\left(1 - \frac{1}{N}\right) \pm 2i\sqrt{L(1-L)}\left(1 - \frac{1}{N}\right) + O\left(\frac{1}{N^2}\right)} \\[2mm]
= \mathring{\pm} (1-\frac{1}{2N})\sqrt{ (1 -2L) \pm 2i\sqrt{L(1-L)}} + O\left(\frac{1}{N^2}\right) \\[2mm]
= \mathring{\pm} (1-\frac{1}{2N})\sqrt{ e^{\pm i2\tau}} + O\left(\frac{1}{N^2}\right) \\[2mm]
= \mathring{\pm} (1-\frac{1}{2N})e^{\pm i\tau} + O\left(\frac{1}{N^2}\right) \\[2mm]
\end{array}$

where $\sin(2\tau) = 2\sqrt{L(1-L)}$.

\vspace{5mm}

These four roots are negatives and/or complex conjugates of each other.  The problem of finding $L=\frac{M}{N}$ can now be reduced to finding the phase of the root of $f(z)$ in the first quadrant, which is made difficult by the fact that $\eta \approx (1-\frac{1}{2N})e^{i\tau}$ lies just inside of the unit circle.  Away from these zeros the frequency response is fairly unrevealing.

\begin{figure}[h!]
\centering
\includegraphics[width=4in]{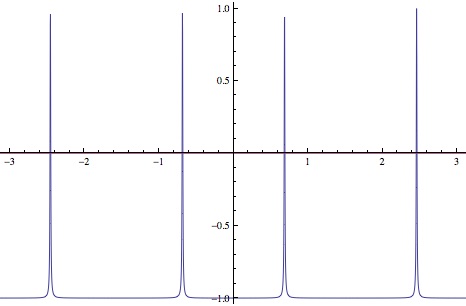}
\caption{$Re\left[R\left(e^{i\theta}\right)\right]$ for $N=100$, $M=40$.}
\end{figure}

With the exception of the thin spikes that occur around the four zeros of $f(z)$, this approximately equal to $-1$, which is the frequency response of a reflection by $-1$ at vertex $c$ (which is approximately what the graph is).

This means that in order to detect one of the $\eta$'s we need to detect an interval in which $S(e^{i\theta}) \ne -1$, and that difference is only detectable within $O\left(\frac{1}{N}\right)$ of the correct value.

So, if a pulse is sent and it is found that $S(e^{i\theta}) \ne -1$, then $2\sqrt{R(1-R)} = \sin(2\theta)$.

\subsection{Example: Complete Graph of Unknown Size}

Here we consider a complete graph with $N+1$ vertices, labeled $A_0, A_1,\ldots,A_N$.  All of these are connected to each other, and $A_0$ is additionally connected to vertex $0$.  We define $${\bf U}|A_j,A_k\rangle = \left(-1+\frac{2}{N}\right)|A_k,A_j\rangle + \frac{2}{N}\sum_{l=0,\,l\ne j}^N |A_k,A_l\rangle$$ and  $${\bf U}|A_j,A_0\rangle = \left(-1+\frac{2}{N+1}\right)|A_0,A_j\rangle + \frac{2}{N+1}\sum_{l=1,\,l\ne j}^N |A_0,A_l\rangle + \frac{2}{N+1}|A_0,0\rangle$$

Define the basis states as:

$\begin{array}{ll}
|\psi_1\rangle = |0,A_0\rangle \\[2mm]
|\psi_2\rangle = \frac{1}{\sqrt{N}}\sum_{j=1}^N |A_0,A_j\rangle \\[2mm]
|\psi_3\rangle = \frac{1}{\sqrt{N(N-1)}}\sum_{j=1}^N\sum_{k=1, k\ne j}^N |A_j,A_k\rangle \\[2mm]
|\psi_4\rangle = \frac{1}{\sqrt{N}}\sum_{j=1}^N |A_j,A_0\rangle \\[2mm]
|\psi_5\rangle = |A_0,0\rangle \\[2mm]
\end{array}$

\vspace{5mm}

We find that ${\bf U}_\alpha$ is:

${\bf U}_\alpha = \left(\begin{array}{ccccc}
0 & 0 & 0 & 0 & \alpha \\[2mm]
\frac{2\sqrt{N}}{N+1} & 0 & 0 & 1-\frac{2}{N+1} & 0 \\[2mm]
0 & \frac{2\sqrt{N-1}}{N} & 1-\frac{2}{N} & 0 & 0 \\[2mm]
0 & -1+\frac{2}{N} & \frac{2\sqrt{N-1}}{N} & 0 & 0 \\[2mm]
-1+\frac{2}{N+1} & 0 & 0 & \frac{2\sqrt{N}}{N+1} & 0
\end{array}\right)$

\vspace{5mm}

It follows that the characteristic polynomial of is

$C(z,\alpha) = z^2\frac{(N^2+N)z^3+(2+N-N^2)z^2+(2-3N+N^2)z+(N-N^2)}{N(N+1)} - \alpha\frac{(N-N^2)z^3+(2-3N+N^2)z^2+(2+N-N^2)z+(N^2+N)}{N(N+1)}$

and therefore (by thm. \ref{important1}) the effective reflection coefficient of $A_0$ is

$$R(z) = z^2S(z) = \frac{(N-N^2)z^3+(2-3N+N^2)z^2+(2+N-N^2)z+(N^2+N)}{(N^2+N)z^3+(2+N-N^2)z^2+(2-3N+N^2)z+(N-N^2)}$$

\vspace{5mm}

Again, $S(z)$ is the coefficient for vertex 0 and $R(z)$ is the coefficient for vertex $A_0$.

Like in the last example this equation couldn't be less clear, but like the last example we can gain some insight by using thm. \ref{window}.  By writing $f(z)$ as

$\begin{array}{ll}
f(z) = z^2\left(z^3 - \left(1-\frac{2}{N}\right)z^2 + \left(1+\frac{2}{N}-\frac{6}{N+1}\right)z - \left(1-\frac{2}{N+1}\right)\right) \\[2mm]
= z^2\left(z^3 - z^2 + z - 1\right) + \frac{z^2}{N}\left(2z^2 - 4z + 2\right) + \frac{z^2}{N^2}\left(6z - 2\right) + O\left(N^{-3}\right) \\[2mm]
= z^2\left(z^2+1\right)\left(z-1\right) + \frac{2z^2}{N}\left(z - 1\right)^2 + \frac{2z^2}{N^2}\left(3z - 1\right) + O\left(N^{-3}\right)
\end{array}$

we can quickly find that to lowest order the roots of $f(z)$ are:

$$\left\{0,0,i-\frac{1+i}{N},-i-\frac{1-i}{N},1-\frac{2}{N^2} \right\}$$

\vspace{5mm}

The last of these is the most difficult to detect and has a constant phase, which means that it is a poor choice for a probing signal.  The other two roots do change phase (a little) and are easier to detect.  We find that, again to first order, $\arg\left(i-\frac{1+i}{N}\right) = \frac{\pi}{2}+\frac{1}{N}$ and $\left|i-\frac{1+i}{N}\right| = 1-\frac{1}{N}$.

\begin{figure}[h!]
\centering
\includegraphics[width=4in]{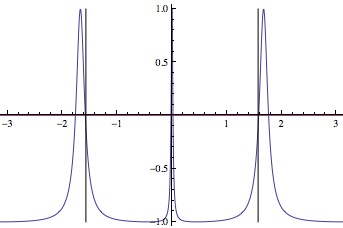}
\caption{$Re\left[R\left(e^{i\theta}\right)\right]$ for $N=10$.  The vertical lines indicate $\theta=\pm \frac{\pi}{2}$.}
\end{figure}

Once again, this effective reflection coefficient of $A_0$ is approximately $-1$ for most values of $\theta$, which is to be expected since $\langle A_0,0|{\bf U}|0,A_0\rangle = -1+\frac{2}{N+1} \approx -1$.  However, if we can find the value of $\theta$ such that $R\left(e^{i\theta}\right)=1$, then we can determine the number of vertices in the complete graph since $R\left(e^{i\theta}\right)=1\Rightarrow \theta=\pm\left(\frac{\pi}{2}+\frac{1}{N}\right)+O(N^{-2})$.

\subsection{Example: A Simple Valve}\label{valve}

Here we have a degree 3 vertex, $D$, connected to three other vertices: $A$, $B$, and $C$.  $D$ is a standard diffusive vertex, so $t=\frac{2}{3}$ and $r=-\frac{1}{3}$, $A$ and $B$ reflect with $0$ and $C$ reflects with some constant $c$.  When attached to a runway vertices $A$ and/or $B$ are vertex zero.

\begin{figure}[h!]
\centering
\includegraphics[width=4in]{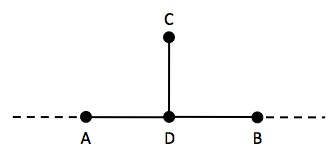}
\caption{The simple valve.  When attached to a runway $A$ and/or $B$ will be vertex 0.}
\end{figure}

\vspace{5mm}

Define the basis states as

$\begin{array}{ll}
|\psi_1\rangle = |A,D\rangle = |in_1\rangle \\[2mm]
|\psi_2\rangle = |B,D\rangle = |in_2\rangle \\[2mm]
|\psi_3\rangle = |C,D\rangle \\[2mm]
|\psi_4\rangle = |D,C\rangle \\[2mm]
|\psi_5\rangle = |D,B\rangle = |out_2\rangle \\[2mm]
|\psi_6\rangle = |D,A\rangle = |out_1\rangle \\[2mm]
\end{array}$

The time step operator, ${\bf U}_0$, is

${\bf U}_0 = \left(\begin{array}{cccccc}
0 & 0 & 0 & 0 & \color{red}{0}_{12} & \color{red}{0}_{11} \\[2mm]
0 & 0 & 0 & 0 & \color{red}{0}_{22} & \color{red}{0}_{21} \\[2mm]
0 & 0 & 0 & c & 0 & 0 \\[2mm]
\frac{2}{3} & \frac{2}{3} & -\frac{1}{3} & 0 & 0 & 0 \\[2mm]
\frac{2}{3} & -\frac{1}{3} & \frac{2}{3} & 0 & 0 & 0 \\[2mm]
-\frac{1}{3} & \frac{2}{3} & \frac{2}{3} & 0 & 0 & 0 \\[2mm]
\end{array}\right)$

\vspace{5mm}

In the above matrix the locations corresponding to the terms that scatter $|in_j\rangle$ into $|out_k\rangle$ are marked in red and labeled $0_{jk}$.  Each of these entries are in fact zero, merely labeled.  Were we using thm \ref{important1} (instead of thm \ref{important2}), then $0_{jk}$ is where $\alpha$ would be inserted in order to calculate $S_{jk}(z)$.

A quick (computer aided) calculation reveals that

$-\left({\bf U}_0-z{\bf I}\right)^{-1} = \left(\begin{array}{cccccc}
\frac{1}{z} & 0 & 0 & 0 & 0 & 0 \\[2mm]
0 & \frac{1}{z} & 0 & 0 & 0 & 0 \\[2mm]
\frac{2c}{z(3z^2+c)} & \frac{2c}{z(3z^2+c)} & \frac{3z}{3z^2+c} & \frac{3c}{3z^2+c} & 0 & 0 \\[2mm]
\frac{2}{3z^2+c} & \frac{2}{3z^2+c} & \frac{-1}{3z^2+c} & \frac{3z}{3z^2+c} & 0 & 0 \\[2mm]
\color{red}{\frac{2(z^2+c)}{z^2(3z^2+c)}} & \color{red}{\frac{-z^2+c}{z^2(3z^2+c)}} & \frac{2}{3z^2+c} & \frac{2c}{z(3z^2+c)} & \frac{1}{z} & 0 \\[2mm]
\color{red}{\frac{-z^2+c}{z^2(3z^2+c)}} & \color{red}{\frac{2(z^2+c)}{z^2(3z^2+c)}} & \frac{2}{3z^2+c} & \frac{2c}{z(3z^2+c)} & 0 & \frac{1}{z} \\[2mm]
\end{array}\right)$

where again, the relevant terms are marked in red.

A subtle point to keep in mind is that these describe $|1,0\rangle$ scattering into $|0,1\rangle$, which are not included in $G$.  To find the effective reflection coefficient of vertex $D$ we simply need to get rid of the delay caused by the two extra edges, and this is done by multiplying by $z^2$ (see fig. \ref{Rdef}).

We see immediately that the effective reflection and transmission coefficients (from either direction) of vertex $D$ are $r(z)=\frac{-z^2+c}{3z^2+c}$ and $t(z)=\frac{2(z^2+c)}{3z^2+c}$ respectively.

When $z^2=c$ we see that $D$ becomes ``transparent", since $r(\pm\sqrt{c})=\frac{-c+c}{3c+c}=0$ and $t(\pm\sqrt{c})=\frac{2(c+c)}{3c+c}=1$.

When $z^2=-c$ we see that $D$ becomes completely reflective, since $r(\pm i\sqrt{c})=\frac{c+c}{-3c+c}=-1$ and $t(\pm i\sqrt{c})=\frac{2(-c+c)}{-3c+c}=0$.

This also provides a tool for determining the value of $c$: if a particular momentum state with eigenvalue $\lambda$ is entirely transmitted, then $c=\lambda^2$.  In fact, this method was used in \cite{farhi} to distinguish between the two results of a boolean function.  When the effective reflection coefficient of a tree graph constructed to execute a particular calculation and attached to vertex $C$ was $-1$ ($1$) the $\pm i$-eigenstate was entirely transmitted (reflected).

\subsection{Example: Square Junction}\label{square}

In this example we look at a graph with four inputs and outputs and discuss what the scattering coefficients mean in terms of eigenstates and signals.  Even with this (16 state) graph we begin to see the use for replacing graphs with a single vertex (with non-trivial scattering coefficients).

\begin{figure}[h!]
\centering
\includegraphics[width=3in]{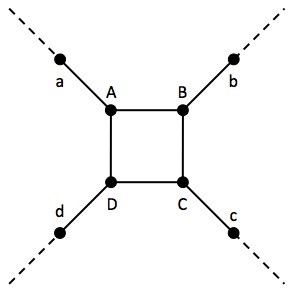}
\caption{A four-way junction.}
\end{figure}

\vspace{5mm}

Define the basis states as

$\begin{array}{ll}
|\psi_1\rangle = |a,A\rangle = |in_1\rangle \\[2mm]
|\psi_2\rangle = |b,B\rangle = |in_2\rangle \\[2mm]
|\psi_3\rangle = |c,C\rangle = |in_3\rangle \\[2mm]
|\psi_4\rangle = |d,D\rangle = |in_4\rangle \\[2mm]
|\psi_5\rangle = |A,B\rangle \\[2mm]
|\psi_6\rangle = |B,C\rangle \\[2mm]
|\psi_7\rangle = |C,D\rangle \\[2mm]
|\psi_8\rangle = |D,A\rangle \\[2mm]
|\psi_9\rangle = |A,D\rangle \\[2mm]
|\psi_{10}\rangle = |D,C\rangle \\[2mm]
|\psi_{11}\rangle = |C,B\rangle \\[2mm]
|\psi_{12}\rangle = |B,A\rangle \\[2mm]
|\psi_{13}\rangle = |D,d\rangle = |out_4\rangle \\[2mm]
|\psi_{14}\rangle = |C,c\rangle = |out_3\rangle \\[2mm]
|\psi_{15}\rangle = |B,b\rangle = |out_2\rangle \\[2mm]
|\psi_{16}\rangle = |A,a\rangle = |out_1\rangle \\[2mm]
\end{array}$

The time step operator, ${\bf U}_0$, is

${\bf U}_0 = \left(\begin{array}{cccc|cccc|cccc|cccc}
0 & 0 & 0 & 0 & 0 & 0 & 0 & 0 & 0 & 0 & 0 & 0 & \color{red}{0_{14}} & \color{red}{0_{13}} & \color{red}{0_{12}} & \color{red}{0_{11}} \\[2mm]
0 & 0 & 0 & 0 & 0 & 0 & 0 & 0 & 0 & 0 & 0 & 0 & \color{red}{0_{24}} & \color{red}{0_{23}} & \color{red}{0_{22}} & \color{red}{0_{21}} \\[2mm]
0 & 0 & 0 & 0 & 0 & 0 & 0 & 0 & 0 & 0 & 0 & 0 & \color{red}{0_{34}} & \color{red}{0_{33}} & \color{red}{0_{32}} & \color{red}{0_{31}} \\[2mm]
0 & 0 & 0 & 0 & 0 & 0 & 0 & 0 & 0 & 0 & 0 & 0 & \color{red}{0_{44}} & \color{red}{0_{43}} & \color{red}{0_{42}} & \color{red}{0_{41}} \\[2mm]\hline
\frac{2}{3} & 0 & 0 & 0 & 0 & 0 & 0 & \frac{2}{3} & 0 & 0 & 0 & -\frac{1}{3} & 0 & 0 & 0 & 0 \\[2mm]
0 & \frac{2}{3} & 0 & 0 & \frac{2}{3} & 0 & 0 & 0 & 0 & 0 & -\frac{1}{3} & 0 & 0 & 0 & 0 & 0 \\[2mm]
0 & 0 & \frac{2}{3} & 0 & 0 & \frac{2}{3} & 0 & 0 & 0 & -\frac{1}{3} & 0 & 0 & 0 & 0 & 0 & 0 \\[2mm]
0 & 0 & 0 & \frac{2}{3} & 0 & 0 & \frac{2}{3} & 0 & -\frac{1}{3} & 0 & 0 & 0 & 0 & 0 & 0 & 0 \\[2mm]\hline

\frac{2}{3} & 0 & 0 & 0 & 0 & 0 & 0 & -\frac{1}{3} & 0 & 0 & 0 & \frac{2}{3} & 0 & 0 & 0 & 0 \\[2mm]
0 & 0 & 0 & \frac{2}{3} & 0 & 0 & -\frac{1}{3} & 0 & \frac{2}{3} & 0 & 0 & 0 & 0 & 0 & 0 & 0 \\[2mm]
0 & 0 & \frac{2}{3} & 0 & 0 & -\frac{1}{3} & 0 & 0 & 0 & \frac{2}{3} & 0 & 0 & 0 & 0 & 0 & 0 \\[2mm]
0 & \frac{2}{3} & 0 & 0 & -\frac{1}{3} & 0 & 0 & 0 & 0 & 0 & \frac{2}{3} & 0 & 0 & 0 & 0 & 0 \\[2mm]\hline

0 & 0 & 0 & -\frac{1}{3} & 0 & 0 & \frac{2}{3} & 0 & \frac{2}{3} & 0 & 0 & 0 & 0 & 0 & 0 & 0 \\[2mm]
0 & 0 & -\frac{1}{3} & 0 & 0 & \frac{2}{3} & 0 & 0 & 0 & \frac{2}{3} & 0 & 0 & 0 & 0 & 0 & 0 \\[2mm]
0 & -\frac{1}{3} & 0 & 0 & \frac{2}{3} & 0 & 0 & 0 & 0 & 0 & \frac{2}{3} & 0 & 0 & 0 & 0 & 0 \\[2mm]
-\frac{1}{3} & 0 & 0 & 0 & 0 & 0 & 0 & \frac{2}{3} & 0 & 0 & 0 & \frac{2}{3} & 0 & 0 & 0 & 0
\end{array}\right)$

\vspace{5mm}

The gridlines here are included to make the matrix more intelligible.  The zeros in the upper-right are labeled according to which input and output they correspond.

To find all of the scattering coefficients we use thm. \ref{important2} and find that the 16 lower left most entries of the negative resolvent are

\begin{equation}\label{squarescatter}
-\left({\bf U}_0-z{\bf I}\right)^{-1} = \left(\begin{array}{cccc|c}
 & & & & \\\hline
\frac{4}{z(9z^2 - 1)} & \frac{16}{27z^4 + 6z^2 - 1} & \frac{4}{z(9z^2 - 1)} & -\frac{9z^4 + 10z^2 - 3}{z^2(27z^4 + 6z^2 - 1)} & \\[2mm]
\frac{16}{27z^4 + 6z^2 - 1} & \frac{4}{z(9z^2 - 1)} & -\frac{9z^4 + 10z^2 - 3}{z^2(27z^4 + 6z^2 - 1)} & \frac{4}{z(9z^2 - 1)} \\[2mm]
\frac{4}{z(9z^2 - 1)} & -\frac{9z^4 + 10z^2 - 3}{z^2(27z^4 + 6z^2 - 1)} & \frac{4}{z(9z^2 - 1)} & \frac{16}{27z^4 + 6z^2 - 1} \\[2mm]
-\frac{9z^4 + 10z^2 - 3}{z^2(27z^4 + 6z^2 - 1)} & \frac{4}{z(9z^2 - 1)} & \frac{16}{27z^4 + 6z^2 - 1} & \frac{4}{z(9z^2 - 1)} \\[2mm]
\end{array}\right)
\end{equation}

\vspace{5mm}

We can gain a little more insight by instead applying thm. \ref{important1} to find $S_{11}(z)$.  First we look at the characteristic polynomial obtained by replacing $0_{11}$ with $\alpha$:

$C_{11}(z) = z^8\left(z^8 - \frac{4}{9}z^6 - \frac{14}{27}z^4 - \frac{4}{81}z^2 + \frac{1}{81}\right) + \alpha z^6\left(\frac{1}{3}z^8 + \frac{4}{27}z^6 - \frac{38}{81}z^4 - \frac{4}{81}z^2 + \frac{1}{27}\right)$

$= \frac{z^6}{81}\left(3z^2+1\right) \left(z^2-1\right) \left[z^2(27z^4 + 6z^2 - 1) + \alpha \left(9z^4 + 10z^2 - 3\right) \right]$

\vspace{5mm}

In this form we see why the scattering coefficients takes such a simple form: in each case $f(z)$ and $g_{jk}(z)$ have several common factors.

The terms that can be factored out of both $f(z)$ and $g_{11}(z)$ correspond to eigenstates that are uninvolved with signals scattering between $|in_1\rangle$ and $|out_1\rangle$.  For example, there are two bound states with eigenvalues $\pm 1$ that aren't involved in any signals (being bound, they have no overlap with any of the input and output states).

Since $S_{11}(z) = -\frac{g_{11}(z)}{f(z)}$, the zeros of $g_{11}(z)$ indicate which signals from $|in_1\rangle$ will return nothing to $|out_1\rangle$.  These zeros are $\lambda = \left\{\pm \frac{1}{3}\sqrt{2\sqrt{13}-5}, \pm \frac{i}{3} \sqrt{2\sqrt{13}+5}\right\}$.  This means that a signal of the form $x[n] = \left(\frac{1}{3}\sqrt{2\sqrt{13}-5}\right)^n$ coming in from any of the inputs will have no reflection; it will scatter entirely out of all of the other outputs.  It's worth pointing out that, in this example, $g_{jk}(z)$ is trivial when $j\ne k$.  As a result {\it every} signal will scatter out of all of the other outputs at least a little (since $g_{jk}(z)$ has no zeros).  For example, there is no signal that can be sent through $|in_2\rangle$ that will not produce an output through $|out_3\rangle$.

Notice also that the real roots are inside of the unit circle and the imaginary roots are outside.  This is a beautiful example of the zeros of $g_{jk}(z)$ neither being determined by the roots of $f(z)$ nor being strictly outside of the unit circle (as they are in the single runway case).

\vspace{5mm}

By using these 3 equations (eq. \ref{squarescatter}) for the square junction we can easily insert this graph into a larger one without adding an additional 16 states, which is important if you want to keep your computational overhead low.

\subsection{Example: The Pruned Tree}

One of the central claims of this paper has been that a subgraph can be replaced with a single vertex with a scattering coefficient we can calculate.  In this example we demonstrate this claim.

We calculate the scattering coefficient of a two-level tree, $G$, by two methods.  First, we calculate it directly using the theorems established in this paper.  Second, by replacing the second level of the tree with a scattering coefficient, $r(z)$, reducing it to a one level tree.

These two methods yield identical results and I argue are in fact equivalent.  That is, replacing a subgraph with it's scattering coefficient is a useful way of simplifying and understanding the calculation of the overall scattering coefficient.

\begin{figure}[h!]
\centering
\includegraphics[keepaspectratio=true, scale = 0.75]{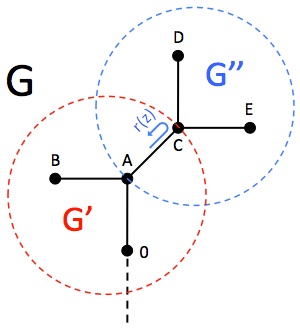}
\caption{``The Pruned Tree".  We can calculate the scattering coefficient of the graph as a whole by replacing $G$ with $G^{\prime}$, and $G^{\prime\prime}$ with the reflection coefficient $r(z)$.}
\end{figure}

We define the basis states as follows

$\begin{array}{ll}
|\psi_1\rangle = |A,0\rangle \\[2mm]
|\psi_2\rangle = |0,A\rangle \\[2mm]
|\psi_3\rangle = |A,B\rangle \\[2mm]
|\psi_4\rangle = |B,A\rangle \\[2mm]\hline
|\psi_5\rangle = |A,C\rangle \\[2mm]
|\psi_6\rangle = |C,A\rangle \\[2mm]\hline
|\psi_7\rangle = |D,C\rangle \\[2mm]
|\psi_8\rangle = |C,D\rangle \\[2mm]
|\psi_9\rangle = |E,C\rangle \\[2mm]
|\psi_{10}\rangle = |C,E\rangle \\[2mm]
\end{array}$

and order them such that the full graph is $G=\{|\psi_1\rangle, \ldots,|\psi_{10}\rangle\}$, the simplified graph is $G^{\prime}=\{|\psi_1\rangle, \ldots,|\psi_6\rangle\}$, and the subgraph to be ``pruned" is $G^{\prime\prime}=\{|\psi_5\rangle, \ldots,|\psi_{10}\rangle\}$.

\vspace{5mm}

First we find the scattering coefficient of the entire graph.

The full time-step operator is

${\bf U} = \left(\begin{array}{cccc|cc|cccc}
0 & -\frac{1}{3} & 0 & \frac{2}{3} & 0 & \frac{2}{3} & 0 & 0 & 0 & 0 \\[2mm]
{\color{red}\alpha} & 0 & 0 & 0 & 0 & 0 & 0 & 0 & 0 & 0 \\[2mm]
0 & \frac{2}{3} & 0 & -\frac{1}{3} & 0 & \frac{2}{3} & 0 & 0 & 0 & 0 \\[2mm]
0 & 0 & 1 & 0 & 0 & 0 & 0 & 0 & 0 & 0 \\[2mm]\hline
0 & \frac{2}{3} & 0 & \frac{2}{3} & 0 & -\frac{1}{3} & 0 & 0 & 0 & 0 \\[2mm]
0 & 0 & 0 & 0 & -\frac{1}{3} & 0 & \frac{2}{3} & 0 & \frac{2}{3} & 0 \\[2mm]\hline
0 & 0 & 0 & 0 & 0 & 0 & 0 & 1 & 0 & 0 \\[2mm]
0 & 0 & 0 & 0 & \frac{2}{3} & 0 & -\frac{1}{3} & 0 & \frac{2}{3} & 0 \\[2mm]
0 & 0 & 0 & 0 & 0 & 0 & 0 & 0 & 0 & -1 \\[2mm]
0 & 0 & 0 & 0 & \frac{2}{3} & 0 & \frac{2}{3} & 0 & -\frac{1}{3} & 0 \\[2mm]
\end{array}\right)$

\vspace{5mm}

The characteristic polynomial is

$C(z) = z^2\left(z^8 +\frac{2}{9}z^6 + \frac{4}{9}z^4 - \frac{2}{9}z^3 + \frac{1}{3}\right)
+ \alpha\left(\frac{1}{3}z^8 - \frac{2}{9}z^6 + \frac{4}{9}z^4 + \frac{2}{9}z^2 + 1\right)$

and according to theorem \ref{important1} the scattering coefficient of vertex $0$ is

$$S(z) = -\frac{3z^8 - 2z^6 + 4z^4 + 2z^2 + 9}{z^2\left(9z^8 + 2z^6 + 4z^4 - 2z^3 + 3\right)}$$

This is the reflection coefficient for the entire graph.  We want the results derived below, where we've ``pruned" the graph, to match this.

\vspace{5mm}

The subgraph uses the basis states $|\psi_5\rangle$ through $|\psi_{10}\rangle$, and has the time-step operator

${\bf U}^{\prime\prime} = \left(\begin{array}{cc|cccc}
0 & {\color{red}\alpha} & 0 & 0 & 0 & 0 \\[2mm]
-\frac{1}{3} & 0 & \frac{2}{3} & 0 & \frac{2}{3} & 0 \\[2mm]\hline
0 & 0 & 0 & 1 & 0 & 0 \\[2mm]
\frac{2}{3} & 0 & -\frac{1}{3} & 0 & \frac{2}{3} & 0 \\[2mm]
0 & 0 & 0 & 0 & 0 & -1 \\[2mm]
\frac{2}{3} & 0 & \frac{2}{3} & 0 & -\frac{1}{3} & 0 \\[2mm]
\end{array}\right)$

The characteristic polynomial of this is

$C^{\prime\prime}(z) = z^2\left( z^4 + \frac{1}{3} \right) + \alpha \left( \frac{1}{3}z^4 +1 \right)$

and again by theorem \ref{important1},

$$S^{\prime\prime} (z) = -\frac{z^4 +3}{z^2\left( 3z^4 + 1 \right)}$$

To find the correct scattering coefficient, the reflection coefficient of vertex $C$, is

$$r(z) = z^2 S^{\prime\prime} (z) = -\frac{z^4 +3}{3z^4 + 1}$$

Again, this $z^2$ is to adjust for the $A, C$ edge (see fig. \ref{Rdef}).

\vspace{5mm}

The ``pruned tree graph", $G^{\prime}$, uses the basis states $|\psi_1\rangle$ through $|\psi_6\rangle$.  We replace the entire subgraph, $G^{\prime\prime}$, with a reflection coefficient $r(z)$ such that ${\bf U}^\prime|A,C\rangle = r(z)|C,A\rangle$.  We can use this when calculating eigenstates, and thus when calculating scattering coefficients.

The pruned tree graph is $G^\prime =\{|\psi_1\rangle, \ldots,|\psi_6\rangle\}$ and its time-step operator is

${\bf U}^\prime = \left(\begin{array}{cccc|cc}
0 & -\frac{1}{3} & 0 & \frac{2}{3} & 0 & \frac{2}{3} \\[2mm]
{\color{red}\alpha} & 0 & 0 & 0 & 0 & 0 \\[2mm]
0 & \frac{2}{3} & 0 & -\frac{1}{3} & 0 & \frac{2}{3} \\[2mm]
0 & 0 & 1 & 0 & 0 & 0 \\[2mm]\hline
0 & \frac{2}{3} & 0 & \frac{2}{3} & 0 & -\frac{1}{3} \\[2mm]
0 & 0 & 0 & 0 & {\color{blue}r(z)} & 0
\end{array}\right)$

\vspace{5mm}

The characteristic polynomial is

$C^\prime(z) = z^2\left( z^4 + \frac{1+r(z)}{3}z^2 - \frac{r(z)}{3} \right) + \alpha \left( \frac{1}{3}z^4 - \frac{1+r(z)}{3}z^2 - r(z) \right)$

The reason this should make sense is that the effective reflection coefficient is the response that graph has as part of an eigenstate.  Indeed, if $\lambda$ is an eigenvalue of the full graph, then it should satisfy $C^\prime(\lambda)=0$.  From theorem \ref{important1} we know that the scattering coefficient of the graph is

$$S^\prime (z) = -\frac{z^4 - (1+r(z))z^2 - 3r(z)}{z^2\left( 3z^4 + (1+r(z))z^2 - r(z) \right)} = -\frac{z^4 - z^2 - \left(z^2 + 3\right)r(z)}{z^2\left( 3z^4 + z^2 + (z^2 - 1)r(z) \right)}$$

If we correctly represented the subgraph, then when we insert $r(z)$ into $S^\prime (z)$ we should recover the full $S(z)$.

$\begin{array}{ll}
S^\prime (z) \\[2mm]
= -\frac{z^4 - z^2 - \left(z^2 + 3\right)r(z)}{z^2\left( 3z^4 + z^2 + (z^2 - 1)r(z) \right)} \\[2mm]
= -\frac{z^4 - z^2 + \left(z^2 + 3\right)\frac{z^4 +3}{3z^4 + 1}}{z^2\left( 3z^4 + z^2 - (z^2 - 1)\frac{z^4 +3}{3z^4 + 1} \right)} \\[2mm]
= -\frac{\left(z^4 - z^2\right)\left(3z^4 + 1\right) + \left(z^2 + 3\right)\left(z^4 +3\right)}{z^2\left( \left(3z^4 + z^2\right)\left(3z^4 + 1\right) - (z^2 - 1)\left(z^4 +3\right) \right)} \\[2mm]

= -\frac{3z^8 - 2z^6 + 4z^4 + 2z^2 + 9}{z^2\left(9z^8 + 2z^6 + 4z^4 - 2z^3 + 3\right)}\\[2mm]
= S(z)
\end{array}$

So, the scattering coefficients obtained from the pruned tree, $G^{\prime}$, and the full tree, $G$, are the same. 

\vspace{5mm}

The four poles of $r(z)$ are of no great concern.  As $S^\prime(z) = S(z)$ for all $z$ for which $r(z)$ is defined, these are removable singularities. 

So, the scattering coefficient of a graph can be found by replacing a subgraph with its own scattering coefficient.  This is useful if, for example, you wanted to do a calculation involving a graph with a lot of identical subgraphs.

\end{document}